\documentclass[lettersize,journal]{IEEEtran}
\usepackage{amsmath,amsfonts}
\usepackage{algorithmic}
\usepackage{algorithm}
\usepackage{array}
\usepackage[caption=false,font=normalsize,labelfont=sf,textfont=sf]{subfig}
\usepackage{textcomp}
\usepackage{stfloats}
\usepackage{url}
\usepackage{verbatim}
\usepackage{graphicx}
\usepackage{cite}
\usepackage{xcolor}
\usepackage{amsthm}
\usepackage{ascmac}
\usepackage{enumitem}
\usepackage{tabularx}
\usepackage{multirow}
\usepackage{etoolbox}
    
\newtheorem{Theorem}{Theorem}
\newtheorem{Definition}{Definition}
\newtheorem{Remark}{Remark}
\newtheorem{Lemma}{Lemma}

\newtheorem{Corollary}{Corollary}

\newcommand{\rev}[1]{\textcolor{black}{#1}}
\newcommand{\revb}[1]{\textcolor{black}{#1}}

\begin{document}

\title{Secret-Key Agreement Using Physical Identifiers for Degraded and Less Noisy Authentication Channels}

\author{Vamoua~Yachongka,~\IEEEmembership{Member,~IEEE,} Hideki~Yagi,~\IEEEmembership{Member,~IEEE,} 
and Hideki~Ochiai,~\IEEEmembership{Fellow,~IEEE}
        % <-this % stops a space
\thanks{V.\ Yachongka and H.\ Ochiai are with the Department of Electrical and Computer Engineering, Yokohama National University, Yokohama, Kanagawa 240-8501 Japan. Corresponding author: Vamoua Yachongka (e-mail: yachongka-vamoua-fs@ynu.ac.jp; hideki@ynu.ac.jp).}
% <-this % stops a space
\thanks{H. Yagi is with the Department of Computer and Network Engineering, The University of Electro-Communications, Chofu, Tokyo, 182-8585 Japan (e-mail: h.yagi@uec.ac.jp).}
\thanks{A part of this paper was presented at the 2022 IEEE Information Theory Workshop (ITW) in \cite{vyo2022itw}. This work was supported in part by the Ministry of Internal Affairs and Communications, Japan, through the contract of ``Research and development on new generation cryptography for secure wireless communication services'' among ``Research and Development for Expansion of Radio Wave Resources,'' under Grant JPJ000254 and by JSPS KAKENHI Grant Numbers JP20K04462 and JP18H01438.}
%\thanks{Manuscript received April 19, 2021; revised August 16, 2021.}
}

%\markboth{IEEE TRANSACTIONS ON INFORMATION FORENSICS AND SECURITY,~Vol.~, No.~,}%


\maketitle

\begin{abstract}
Secret-key agreement using physical identifiers is a promising security protocol for the authentication of users and devices with small chips, owing to its lightweight security. In the previous studies, the fundamental limits of such {a protocol} were analyzed, and 
%In this study, we investigate the fundamental limits of such model key Kittichokechai and Caire (2016) investigated the optimal trade-off in a secret-key agreement model with physical identifiers, where the structure of the authentication channels {is} similar to the wiretap channels, {from information theoretic approaches}. Later, the model was extended by G\"unl\"u et al.\ (2018) introducing noise in the enrollment phase and cost-constrained {actions} at the decoder.
the results showed that two auxiliary random variables were involved in the capacity region expressions. However, with two auxiliary random variables,
%the complexity of calculating the capacity region may be prohibitively high. 
\revb{it is difficult to directly apply the expressions to derive the computable forms of the capacity regions for certain information sources such as binary and Gaussian sources, which hold importance in practical applications. In this paper, we explore the structure of authentication channels and reveal that for the classes of degraded and less noisy authentication channels, a single auxiliary random variable is sufficient to express the capacity regions. As specific examples, we use the expressions with one auxiliary random variable to derive the computable forms for binary and Gaussian sources}. Numerical calculations for the Gaussian case show the trade-off between secret-key and privacy-leakage rates under a given storage rate, which illustrates how the noise in the enrollment phase affects the capacity region.
\end{abstract}


\section{Introduction}
%This sort of security is based on intrinsic behaviors of the device or users, so there is no need of powerful processors to implement like public key cryptography. This advantage underlies the potential of applying this technique for device authentications in the 5th/6th communication generation.

\IEEEPARstart{I}{n} the age of fast and momentous advancements in communication technologies, the number of Internet-of-Things (IoT) devices {has} increased remarkably. Since IoT devices equipped with small chips have resource-constrained capabilities, they may not be suitable for deploying high-profile cryptography {schemes} such as public-key encryption/decryption for device {authentication}. Lightweight security protocols handily feasible on physical layers have been receiving recent attention to a greater extent since they enable the devices to securely communicate with low latency as well as low power consumption \cite{blochetal2021}.
%Recently, possibilities of deploying this sort of technology in upcoming sixth-generation communication systems have been actively discussed in the literature, e.g., \cite{parambage}, \cite{Nguyenetal}.

Secret-key agreement in which physical identifiers are used as information sources to generate secret keys for authentication, called {\em authentication system} in this paper, has emerged as a promising candidate {since} it provides a low-complexity design, consumes less power, and preserves secrecy \cite{herder2014}. As authentication can be performed on demand, the cost is lower than that of key storage in non-volatile random access memories \cite{pappu2001}, \cite{puf2003}. %Typical examples of biometric identifiers include fingerprints, irises, and faces \cite{jian2009}, whereas
Physical identifiers could be physical unclonable functions (PUFs), making use of intrinsic manufacturing variations of the integrated circuit to produce source sequences \cite{bohm2012}. \revb{Several PUF designs have been proposed over the last few decades and can be largely classified into either strong PUFs or weak PUFs. We focus on weak PUFs such as static random-access memory (SRAM) PUFs and ring oscillator (RO) PUFs since they produce reliable challenge-response pairs that can be used as unique cryptographic keys for IoT device security \cite{zhangetal2022}}.
%These identifiers can be used as information sources to generate secret keys for authentications.
%Biometric data belong to human beings from birth, however, PUF is a function embedded inside devices and it makes use of the randomness property of intrinsic manufacturing variations of the integrated circuit to produce source sequences \cite{bohm2012}. %Some well-known PUFs include, but not limited to ring oscillator PUF, static random access memory PUF, arbiter PUF, and so forth \cite{herder2014}.
Although generating processes are different, PUFs and biometric identifiers have several aspects in common, and nearly all assumptions and analyses of PUFs can be applied to biometric identifiers \cite{gunlue2020}. Thus, the theoretical results developed in this study should be applicable to the scenario where biometric identifiers are treated as sources.

\begin{figure}[!t]
    \centering
    \includegraphics[scale=0.45]{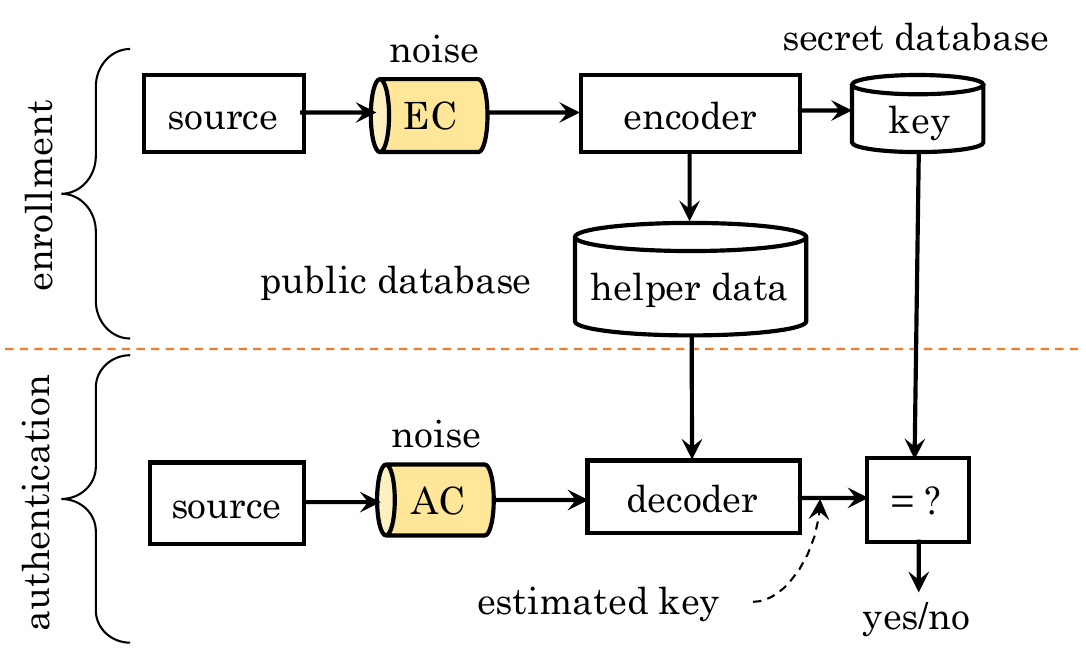}
    \vspace{-3mm}
    \caption{A basic concept of secret-key agreement using physical and biometric identifiers \cite{itw3}.}
    \vspace{-5mm}
    \label{model-intro}
\end{figure}

%\revb{Several PUF designs have been proposed over the last few decades and can be largely classified into either strong PUFs or weak PUFs. We focus on weak PUFs such as SRAM PUFs and ring oscillator PUFs (RO PUFs) since they produce reliable challenge-response pairs that can be used as unique cryptographic keys for IoT device security \cite{zhangetal2022}}.
A block diagram related to the data flows of an authentication system with PUFs is illustrated in Figure\ \ref{model-intro}, and the system consists of two phases, i.e., enrollment (top) and authentication (bottom) phases. In the enrollment phase, observing a measurement of the source sequence via a channel, which is assumed to be noise-free in some previous studies, the encoder generates a pair of {\em secret key} and {\em helper data}. The helper data is shared with the decoder via a noiseless public channel to assist in the reconstruction of the secret key\footnote{It is assumed that the secret key is stored in a secure database whose location is unknown to an eavesdropper; however, the eavesdropper eavesdrops on the helper data from the public database, which can be thought of as a public channel connecting the encoder and decoder, and utilizes it to examine the statistical behavior of the secret key.}. In the authentication phase, the decoder estimates the secret key using the helper data and another measurement observed through a channel in this phase \cite{itw3}, \cite{lhp}. In this paper, the channels in the enrollment and the authentication phases are called the {\em enrollment channel} (EC) and {\em authentication channel} (AC), respectively. EC and AC are modeled to represent the noises added to the identifiers during the enrollment and authentication phases, respectively.

Relevant practical applications of the system described above include biometrics-based access control systems \cite{rane2016}, fuzzy extractor schemes \cite{dodis2008}, \cite{romero2023}, and field-programmable gate array (FPGA) based key generation with PUFs for IoT device authentication \cite{ANANDAKUMAR2021}. As a connection to physical layer security, PUFs are deployed to assist with key generation in poor scattering environments to enhance the randomness of bit sequences extracted from wireless channels, and it has been demonstrated that a higher secret-key generation rate is realizable \cite{Tasneem2023}.

\subsection{Related Work}
Seminal studies \cite{itw3} and \cite{lhp} independently investigated {the fundamental limits of secret-key and privacy-leakage rates, called the {\em capacity region}}, of the authentication systems. The capacity region elucidates the best possible trade-off between secret-key and privacy-leakage rates. The revealed trade-off may provide direct insights and serves as significant indicators for researchers seeking to design good practical codes that could achieve the largest achievable secret-key rate and the lowest implementable privacy-leakage rate for an authentication system\footnote{Note that when referring to the capacity region in the upcoming sections, it includes an extra dimension, the storage rate, along with the secret-key and privacy-leakage rates. To decrease memory usage in the public database, the storage rate should be minimized, similar to the privacy-leakage rate.}.
%Particularly,
%Indeed, the models of these works can be viewed as the well-know secret-key agreement problem (forward transmission only) discussed in \cite{ac1993}, \cite{maurer1993} with adding a privacy constraint.
In \cite{itw3}, eight different systems were taken into consideration, but among them, the generated-secret (GS) and chosen-secret (CS) models are {the} two major systems that are closely related to real-life applications and have been frequently analyzed in subsequent studies\footnote{The difference between the two models appears in the enrollment phases. In the GS model, the secret key is extracted from the measurement of identifiers observed at the encoder and does not need to be saved in the public database. By contrast, in the CS model, the secret key, chosen uniformly and independently of other random variables, is combined with the measurement.
%The information related to the secret key is also
The combined information, which contains data relevant to the secret key but not the plain form of the key, is stored in the database so that the decoder can reliably estimate the secret key. Hence, compared with the GS model, the minimum amount of storage rate required for the CS model is larger in general. See \cite[Section III]{itw3} for a more comprehensive explanation.}.

The secret-key capacity increases for multiple rounds of enrollments and authentications in the GS model \cite{kuster2019} and CS model \cite{kusters2021} with static random-access memory PUFs (SRAM PUFs). The work \cite{itw3} is extended to include a storage constraint \cite{KY}, {a} multi-identifier scenario with joint and distributed encoding \cite{chou2019}, and polar codes for achieving the fundamental limits \cite{chou2015}. The theoretical results in \cite{itw3}--\hspace{-0.01mm}\cite{chou2015} are clarified under a common assumption, i.e., the EC is noiseless, and this particular model is known as a {visible source model}. Recently, the capacity regions of the GS and CS models have been characterized in a more realistic setting where EC is noisy \cite{gunlu2018}, and this model is called a {hidden source model}. As an extended scenario of authentication systems, the GS and CS models that involve not only secret-key authentication but also user identification can be found in, e.g., \cite{itw2015,vy2020,zhou2022}.

For practical code constructions on the authentication systems, some state-of-the-art approaches for binary source sequences are investigated in \cite{chen2019} for polar codes and in \cite{2019gunlu} for both Wyner-Ziv and nested polar codes. Compared to the simulation results in \cite{chen2019} and \cite{2019gunlu}, better performance in terms of secret-key versus storage ratio is achieved by deploying nested randomized polar subcodes \cite{2021gunlu}. Lately, a model with non-binary sources is developed in \cite{muelich2021} with multilevel coding, and its performance is also evaluated by taking coded modulation and shaping techniques into consideration \cite{fisher2022}.

The capacity region of a GS model with the structure of AC following the channel of the wiretap channels or two-receiver broadcast channels with confidential messages \cite{csiszar1978} was investigated in \cite{kc2015}. In this model, AC is composed of the channel to the encoder, referred to as the main channel, and the channel to the eavesdropper (Eve), referred to as Eve's channel. Eve can obtain not only the helper data transmitted over public channels but also a correlated sequence of the source identifiers via her channel. This setup can be viewed as the source model of key-agreement problems \cite{ac1993,cn2000,wataoha2010} with one-way communication only and a privacy constraint. The privacy constraint is imposed to minimize the information leakage of the identifiers, and in general, its analysis becomes challenging especially when the noise in the enrollment phase is taken into account \cite{gunlu2018}. An extension of the work \cite{kc2015} by considering noisy EC and action cost at the decoder was presented in \cite{gksc2018},
%. More precisely, while the capacity regions of the GS model with passive and active Eve were studied in \cite{kc2015}, the work \cite{gksc2018} focused only on passive Eve but gave closed-form expressions of the capacity regions for both GS and CS models,
and in both \cite{kc2015} and \cite{gksc2018}, it was shown that the resulting expressions of the capacity regions involve two auxiliary random variables for a general class of ACs.

In a different setting, the GS and CS models with joint-measurement channels, where EC and AC are modeled as broadcast channels \cite{GK} to assume correlated noises in the measurements, were {examined} in \cite{gsk2019}. Models with joint-measurement channels that incorporate Eve's channel can be found in \cite{gs2021}. These studies analyzed the capacity regions for some classes of broadcast channels, e.g., degraded and less noisy channels \cite{GK}. In a similar manner, we also investigate the capacity regions of the authentication systems for similar classes of channels, but the models and the point to which we direct our attention are different from those of \cite{gsk2019}, \cite{gs2021}. More precisely, we deal with the models with separate measurements as in \cite{gksc2018}, and focus on the structure of AC, e.g., the main channel is less noisy than Eve's channel or Eve's channel is degraded {with respect to} the main channel, to simplify the expressions of the capacity regions with two auxiliary random variables that have been characterized in the paper.

\subsection{Motivations}

%Several PUF designs have been proposed over the last few decades and can be largely classified into either strong PUFs or weak PUFs. We focus on weak PUFs such as SRAM PUFs and ring oscillator PUFs (RO PUFs) since they produce reliable challenge-response pairs that can be used as unique cryptographic keys for IoT device security \cite{zhangetal2022}.
\revb{In real-life applications, the observations of PUFs and biometric identifiers are usually corrupted by noise. For instance, the measurements of PUFs' signals are affected by surrounding environments of integrated circuits such as temperature variation, change of supply voltage, and electronic noise \cite{herder2014}, \cite{gunlue2020}. Likewise, a scanned picture of a fingerprint corresponds to a noisy version of its original image. Therefore, the assumption of the hidden source model as in \cite{gksc2018} is considered to be a more realistic setting compared to that of the visible source model \cite{kc2015}. We thus adopt the setting of \cite{gksc2018} on our model.}

%\rev{The hidden model source model \cite{gksc2018}, in which the noise in the enrollment phase is assumed, is applied to the system model considered in this paper. This is a  because }
%Therefore, considering {hidden source model} as in \cite{gksc2018} is more realistic. %{Indeed, we focus on this setting and our model is the same as the one analyzed in \cite{gksc2018}.}
%In \cite{gksc2018}, it is shown that two auxiliary {random variables} ({random variable}s) appear in the capacity region expressions for a general class of ACs.

As we mentioned in the previous subsection, the expressions of the capacity regions of the GS and CS models characterized in \cite{gksc2018} under a general class of AC involve two auxiliary random variables. \revb{Nevertheless, these expressions are impractical for developing the computable and tight bounds for some specific information sources and channels directly.
%, which are often discussed in practice.
%due to the difficulty of handling two auxiliary random variables.
%which are often discussed in practice}.
Therefore, we explore and identify the classes of ACs that require only one auxiliary random variable for expressing the capacity regions, and use the simplified expressions to derive the computable forms for those specific sources and channels.}
%\revb{In addition to reducing complexity, the simplified expression may lead to characterizing the tight bound on some specific types of information sources, which is important from a practical viewpoint.}

In this paper, we first investigate and characterize the capacity region with a single auxiliary random variable of the authentication systems for discrete sources and then apply this result to derive the capacity regions of GS and CS models for binary sources and channels. As an application of the systems with binary sources, it is well-known that SRAM-PUF responses are binary, and the outputs of sources and channels of SRAM PUFs can be modeled as binary bit sequences \cite{kuster2019}.

Furthermore, the measurements of the majority of PUFs are represented by continuous values. %\rev{See also \cite{enrique2019} for an understandable description}.
As an instance, the samples generated by RO PUFs obey a Gaussian distribution \cite{Gebali2022}. In addition, the noise in most communication channels is modeled as additive white Gaussian noise (AWGN). Motivated by this nature, we later extend {the} GS and CS models considered in \cite{gksc2018} to characterize the capacity regions for Gaussian sources and channels.

\subsection{Summary of Contributions}

%In this study, we are interested in exploring classes of authentication channels that need only one auxiliary {random variable} in the characterization of the capacity region of characterizations.
%For simplicity reasons, we do not assume the correlated noise in the measurement channels
%In \cite{gsk2019}, the fundamental limits of the authentication systems, when the broadcast channels are degraded and less noisy channels, were investigated.
Unlike the technique used in \cite{kc2015}, \cite{gksc2018}, we apply information-spectrum methods \cite{wataoha2010}, \cite{han2003} to derive our main results. An advantage of leveraging {these methods} is that the argument does not depend on the size of {the} source alphabet, so it can also {encompass} continuous sources. The main contributions of this work are listed as follows:
\begin{itemize}
    \item We demonstrate that one auxiliary random variable suffices to characterize the capacity regions of the GS and CS models when ACs are in the class of less noisy channels. Though less noisy ACs are a subclass of a general class of ACs, our results are not obtainable by a trivial reduction from the result derived in \cite{gksc2018} under the general class of ACs.
    %, which will be seen later in the proof {of Theorem \ref{th2}}.
    %\item We show that as long as the {AC is} in {the} class of less noisy {channels}, one auxiliary random variable suffices to characterize the capacity regions among secret-key, storage, and privacy-leakage rates for {the} GS and CS models. Though less noisy ACs are a subclass of general {class of} ACs, the characterized regions are not obtainable by a trivial reduction {from} the {characterization for} general {class of} ACs derived in \cite{gksc2018}, which we will see later in the proof {of Theorem \ref{th2}}.
    %To deal with the difficulty in the analysis of the privacy-leakage rate, we give an alternative proof of \cite[Lemma 4]{kitti2015} based on information-spectrum methods.
    \item We apply the simplified expressions to derive the capacity regions for binary sources under less noisy ACs, which is a more general setting than the one {discussed} in \cite[Section IV]{gksc2018}. To obtain the tight regions, we establish a new lemma and use it to match the inner and outer bounds.
    \item The work \cite{vyo2022} is extended to characterize the closed-form expressions of the capacity regions for a {hidden source model}. Also, numerical calculations of the Gaussian case are provided to demonstrate the trade-off between secret-key and privacy-leakage rates in the visible and {hidden source models} and to capture the effects of noise in the enrollment phase toward the capacity region.
\end{itemize}

\subsection{Modeling Assumptions}
We assume that each symbol in the source sequences is independently and identically distributed (i.i.d.). Techniques such as principal component analysis \cite{kelkboom2010} and transform-coding-based algorithms \cite{gunluwifs2015} can be applied to convert biometric and physical identifiers into a vector having (nearly) independent components. However, under various environments and conditions, it may not be feasible to completely remove the correlations among symbols in the source sequence. For simplicity in the analysis, in this paper, we derive all the results under the assumption that every symbol of the source and measurement vectors is i.i.d.~and generated according to a joint distribution.

\revb{In principle, Eve can be classified as either a passive or active eavesdropper. In this paper, we only focus on a passive attack and do not address the issues of active attacks on PUFs, e.g., machine learning and side-channel attacks \cite[Section IV]{gksc2018}. The obtained results are analyzed under the common assumption that a PUF is capable of fending off these invasive attacks that may transform the physical features of PUF outputs permanently \cite{gunlue2020}.}
%Therefore, under this assumption, the remaining possible attack is from a passive Eve, and in the current paper, we concentrate on evaluating the security and leakage of such an attack scenario.}

\begin{table}[!t]
   \centering
   \caption{List of notation}
   \label{tab:notation}
   \begin{tabularx}{\columnwidth}{|c|X|}
     \hline
     Notation & Descriptions \\ \hline
     $n$ & Block length \\ \hline
     $\log x$ & The natural logarithm for $x > 0$  \\ \hline
     $[k:t]$ & Set $\{k,\cdots,t\}$ for integers $k$ and $t$ such that $k < t$  \\ \hline
     $[p,q]$ & The closed interval from $p$ to $q$ for $p,q \in \mathbb{R}$  \\ \hline
     \multirow{2}*{$p*q$} & Convolution operator, defined as $p*q=p(1-q) + (1-p)q$ for $p \in [0,1]$ and $q \in [0,1]$ \\ \hline
     \multirow{2}*{$A_k^t$} & Partial sequence of $A^n$, i.e., $A^t_k = (A_k,\cdots,A_t)$, for any $[k:t] \subseteq [1:n]$ \\ \hline
     %$R_S,R_J,R_L$ & Secret-key rate, storage rate, privacy-leakage rate \\ \hline
    %$\mathcal{R}_G,~\mathcal{R}_C$ & The capacity regions of CS and GS models \\ \hline
    $I(A;B)$ & Mutual information of random variables $A$ and $B$ \\ \hline
    \multirow{2}*{$A-B-C$} & Markov chain, implying that random variables $A$ and $C$ are conditionally independent given $B$ \\
    \hline
    $\mathcal{A},~|\mathcal{A}|$ & A set $\mathcal{A}$,~ cardinality of the set $\mathcal{A}$ \\
    \hline
    \multirow{2}*{$\beta$} & Parameter used in taking union on the capacity region for binary sources \\ \hline
    \multirow{2}*{$\alpha$} & Parameter used in taking union on the capacity region for Gaussian sources \\ \hline
    $\rho$ & Correlation coefficient between two random variables \\ \hline
    $\delta$, $\gamma$ & Small enough positive numbers \\ \hline
    $\rm{Bern}(0.5)$ &  Bernoulli distribution with outcome probability $0.5$ \\ \hline
    $\mathcal{N}(0,\sigma^2)$ & Gaussian distribution with zero mean and variance $\sigma^2$ \\ \hline
   \end{tabularx}
\vspace{-5mm}
\end{table}

\subsection{Notation and Organization}
Italic uppercase $A$ and lowercase $a$ denote a {random variable} and its realization, respectively. $A^n = (A_{1},\cdots ,A_{n})$ represents a string of {random variables} and subscript represents the position of a random variable in the string. $P_A(\cdot)$ denotes the probability mass function of the random variable $A$. $H(\cdot)$ and $H_b(\cdot)$ denote the Shannon entropy and the binary entropy function, respectively. For other notation, refer to Table \ref{tab:notation}.

The rest of this paper is organized as follows: In Section \ref{sect2}, we introduce the system models and formulate achievability definitions. Section \ref{sect3} derives the capacity regions of the authentication systems with one auxiliary random variable, and Section \ref{sect4} focuses on binary and Gaussian examples.
%The proof of our main results is available in appendixes.
Finally, concluding remarks and future work are given in Section \ref{sect5}.

%Secret key-based authentication (SKA) systems are generally designed to perform private authentication of users based on secret keys, usually generated from biometric identifiers \cite{jian2009} or physical unclonable functions \cite{bohm2012}. In recent years, there has been a bunch of literature focusing on investigating the fundamental limits of SKA systems from information-theoretic perspectives. {In} the analysis of the {SKA} systems, {a new condition} called privacy constraint is {added to} the problem formulations of the well-known secret-key agreement {(the source type model with forward communication only)} discussed in, e.g., \cite{ac1993}--\hspace{-0.1mm}\cite{wataoha2010}. Therefore, many existing tools used for solving the key agreement problems are quite useful {to characterize} the capacity regions of the {SKA} systems as well. 

%tangible

\section{System Models and Problem Formulations} \label{sect2}
\subsection{System Models}
\begin{figure}[!t]
    \centering
    \vspace{-1mm}
    \includegraphics[scale=0.55]{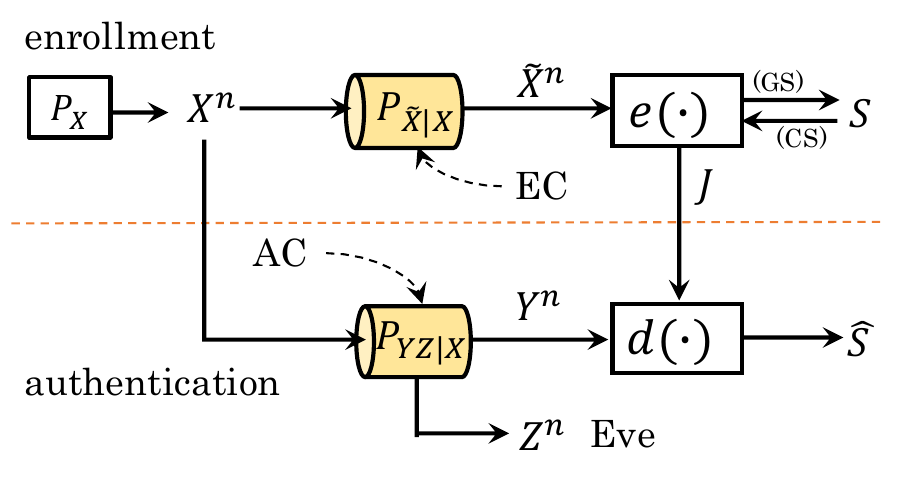}
    \vspace{-6mm}
    \caption{System models in the presence of Eve: The arrows attached with (GS) and (CS) denote the directions of the secret keys in the {GS} and {CS} models, respectively.}
    \label{fig:model}
    \vspace{-4mm}
\end{figure}

The GS and CS models, with mathematical notations, are depicted in {Figure} \ref{fig:model}. %{Arrows} (GS) and (CS) indicate the directions of the secret key of the former and {latter} models.
%Phases (I) and (I\hspace{-0.1mm}I) indicate the enrollment and authentication phases, respectively. %In the former model, the secret key is extracted from measurement, while in the {latter} one, it is chosen independently and {bonded} into the measurements.
The sequences $(\Tilde{X}^n,X^n,Y^n,Z^n)$ are i.i.d., and their joint distribution is factorized as
$
    P_{\Tilde{X}^nX^nY^nZ^n} %&= P_{\Tilde{X}^n|X^n}\cdot P_{X^n} \cdot P_{Y^nZ^n|X^n} \nonumber \\
    = \prod_{t=1}^n P_{\Tilde{X}_t|X_t}\cdot P_{X_t} \cdot P_{Y_tZ_t|X_t}. %\label{jointd}
$

Let $\mathcal{S}_n = [1:M_S]$ and $\mathcal{J}_n = [1:M_J]$ be the sets of secret keys and helper data, respectively. {Here, $M_S$ and $M_J$ stand for the largest values in the sets from which secret key and helper data take values}. The random vectors $\Tilde{X}^n$ and $(Y^n,Z^n)$ denote the measurements of the identifier $X^n$, generated from i.i.d. source $P_{X}$, via EC $(\mathcal{X},P_{\Tilde{X}|X}, \Tilde{\mathcal{X}})$ and {AC} $(\mathcal{X},P_{YZ|X}, \mathcal{Y}\times\mathcal{Z})$, respectively. Assume that all alphabets $\Tilde{\mathcal{X}}$, $\mathcal{X}$, $\mathcal{Y}$, and $\mathcal{Z}$ are finite, but this assumption will be relaxed in {Section \ref{gauss-source}}.

In {the} GS model, observing the measurement $\Tilde{X}^n$, the encoder $e$ generates a helper data $J \in \mathcal{J}_n$ and a secret key $S \in \mathcal{S}_n$; $(J,S) = e(\Tilde{X}^n)$. The helper data $J$ is shared with the decoder via a noiseless public channel. Detecting $Y^n$, the decoder $d$ estimates the secret key generated at the encoder using $Y^n$ and helper data $J$; $\widehat{S}=d({Y^n},J)$, {where $\widehat{S}$ denotes an estimation of the secret key $S$}. In the CS model, the secret key $S$ is chosen uniformly from $\mathcal{S}_n$ and {is} independent of other {random variables}. It is embedded into the measurement $\Tilde{X}^n$ to form the helper data $J$; $J = e(\Tilde{X}^n,S)$. For the decoder, similar to the decoder of {the} GS model, the estimate {is produced as} $\widehat{S} = d(Y^n,J)$.

As the helper data $J$ is sent over public channels, Eve can completely eavesdrop on this information. In addition to the helper data, Eve has a sequence $Z^n$, an output of the marginal channel $P_{Z|X}$, and both $J$ and $Z^n$ are exerted to learn the secret key $S$ as well as the source identifier $X^n$. In essence, the information leaked to Eve regarding the identifier can not be made negligible because of the high correlation among $X^n$, $J$, and $Z^n$. However, it is possible to decelerate the distributions of $S$ and $(J,Z^n)$ and make them almost independent, so Eve may be able to recover only some insignificant bits but not the entire secret key based on the data available on her side.

\vspace{-3mm}
\subsection{Problem Formulations for {the} GS and CS Models}
In this section, the formal achievability definitions of {the} GS and CS models are provided. We begin with {the} GS model.
\begin{Definition} \label{def1}
A tuple of secret-key, storage, and privacy-leakage rates $(R_S,R_J,R_L)\in \mathbb{R}^3_+$ is said to be achievable for {the} GS model if for sufficiently small $\delta > 0$ and large enough $n$ there exist pairs of encoders and decoders satisfying
\begin{align}
\Pr\{\widehat{S} \neq S\} &\leq  \delta,~~~~~~~~~~~~~~~~~~~\mathrm{(error ~probability)}\label{errorp} \\
H(S) + {n\delta} &\ge \log M_S \ge  n(R_S - \delta),~~~\mathrm{(secret\text{-}key)} \label{secretk} \\
\log{M_J} &\leq n(R_J + \delta),~~~~~~~~~~~~~~~~~~~\mathrm{(storage)} \label{storage} \\
I(S;J,Z^n) &\leq \delta,~~~~~~~~~~~~~~~~~~~~~\mathrm{(secrecy\text{-}leakage)} \label{secrecy} \\
I(X^n;J,Z^n) &\leq n(R_L + \delta).~~~~~~~~~~\mathrm{(privacy\text{-}leakage)} \label{privacy}
\end{align}
Also, $\mathcal{R}_G$ is defined as the closure of the set of all achievable rate tuples for {the} GS model, called the capacity region.
\qed
\end{Definition}

The technical meaning of each constraint in Definition \ref{def1} can be interpreted as follows: Condition \eqref{errorp} evaluates the error probability of estimating the secret key. This is related to the reliability of the authentication systems and the probability must be bounded by a sufficiently small number $\delta$. Equation \eqref{secretk} is the constraint on the secret-key rate, and the generated key should be forced to be nearly uniform in the entropy sense so as to extract as large a key size as possible. Constraint \eqref{storage} is imposed to minimize the size of the local random codebook that is required for enrollment and authentication. The rate of the codebook must not exceed a given storage rate $R_J$.

Equation \eqref{secrecy} measures the information leaked about the secret key to Eve, called secrecy leakage, and the secrecy leakage is evaluated under a strong secrecy criterion, which requires that the amount of leakage should be bounded by a small value regardless of the block length $n$. In other words, Eve can only obtain an ignorable amount of information regarding the secret key through the helper data and the correlated sequence. The last condition \eqref{privacy} assesses the amount of privacy leakage for the biometric or physical identifiers to Eve. In general, unlike the secrecy leakage \eqref{secrecy}, it is infeasible to make this amount vanish since the helper data itself are generated from $\Tilde{X}^n$, a correlated sequence of $X^n$, and $Z^n$ is also correlated to $X^n$. However, it is important to minimize this quantity to protect the sensitive data of users or the characteristics of PUFs embedded inside the integrated circuits of IoT devices.

The achievability definition of CS model is defined below.
\begin{Definition} \label{def2}
A tuple of $(R_S,R_J,R_L)\in \mathbb{R}^3_+$ is said to be achievable for {the} CS model if for any $\delta > 0$ and large enough $n$ there exist pairs of encoders and decoders satisfying all the requirements imposed in Definition \ref{def1} {with replacing \eqref{secretk} by
%\vspace{-1mm}
\begin{align}
    \log M_S \ge n(R_S - \delta). \label{cs-keycon}
\end{align}}
We define $\mathcal{R}_C$ as the capacity region of {the} CS model.
\qed
\end{Definition}

The interpretations of the constraints in Definition \ref{def2} are the same as that in Definition \ref{def1}; therefore, the details are omitted. In \eqref{cs-keycon}, the enforcement of the secret key to be uniform is no longer needed for the CS model as the secret key is uniformly chosen from the set $\mathcal{S}_n$.

There are other possible ways to define the secrecy-leakage and privacy-leakage in the authentication systems such as conditional entropy and variational distance. Nevertheless, in this paper, we adopt mutual information as the main metric, as in \cite{kc2015} and \cite{gksc2018}, so that it would be easy for us to connect our main results with those clarified in previous studies.
\vspace{-2mm}
\subsection{The Capacity Regions With Two Auxiliary {Random Variables}}

To facilitate the understanding of our main {contributions} in Section \ref{sect3}, we highlight a complete characterization of the capacity regions of {the} GS and CS models (without action costs) derived in \cite{gksc2018} for discrete sources.
\begin{Theorem} (G\"unl\"u et al. \cite[Theorems 3 and 4]{gksc2018}) \label{gunlu2018}
The capacity regions of {the} GS and CS models under {the} general class of ACs are given by
\vspace{-1mm}
\begin{align}
{\mathcal{R}_G} = &\bigcup_{P_{U|\Tilde{X}},P_{V|U}}\Big\{(R_S,R_J,R_L)\in \mathbb{R}^3_+: \nonumber \\
&R_S \leq I(Y;U|V)-I(Z;U|V),~~~R_J \geq I(\Tilde{X};U|Y), \nonumber \\
&R_L \geq I(X;U,Y) - I(X;Y|V) + I(X;Z|V)\Big\}, \label{gunlu333} \\
{\mathcal{R}_C} = &\bigcup_{P_{U|\Tilde{X}},P_{V|U}}\Big\{(R_S,R_J,R_L)\in \mathbb{R}^3_+: \nonumber \\
&R_S \leq I(Y;U|V)-I(Z;U|V),\nonumber \\
&R_J \geq I(\Tilde{X};U|Y) + I(Y;U|V)-I(Z;U|V), \nonumber \\
&R_L \geq I(X;U,Y) - I(X;Y|V) + I(X;Z|V)\Big\}, \label{gunlu333cs}
\end{align}
where auxiliary {random variables} $U$ and $V$ {satisfy} the Markov {chain} $V-U-\Tilde{X}-X-(Y,Z)$ and their {cardinalities} are limited to $|\mathcal{V}| \le |\Tilde{\mathcal{X}}| + 6$ and $|\mathcal{U}| \le (|\Tilde{\mathcal{X}}| + 6)(|\Tilde{\mathcal{X}}| + 5)$.
\qed
\end{Theorem}

The single-letter expressions of the regions above associate two auxiliary random variables $U$ and $V$. Theorem \ref{gunlu2018} tells us that similar to the conclusion drawn in \cite{cn2000} for the key-agreement problem, two auxiliary random variables are required for expressing the capacity regions of the authentication systems for the general class of ACs.

In general, once the single-letter expressions for discrete sources are established, it is common to characterize the capacity region for special cases via such expressions. However, it is challenging to directly employ the expressions in \eqref{gunlu333} and \eqref{gunlu333cs} so as to characterize the capacity regions for binary and Gaussian sources due to the difficulty of handling two auxiliary random variables.
%For instance, the capacity regions for binary sources derived through \eqref{gunlu333} and \eqref{gunlu333cs} is not yet tight, and this is still an open problem.}
%To calculate these regions, we have to operate the transition probabilities of both test channels\footnote{{These channels are used to calculate the mutual information that appears on the right-hand side of each rate constraint in \eqref{gunlu333} and \eqref{gunlu333cs}, and thus, they have a direct impact on the capacity regions of both the GS and CS models.}} $P_{U|\Tilde{X}}$ and $P_{V|U}$ together, which may be computationally infeasible.
%which requires more time and space complexity compared to handling only one test channel.
In the next section, we explore the classes of ACs such that the capacity regions can be expressed by one auxiliary random variable.

\section{Statement of Main Results} \label{sect3}
As mentioned in the introduction, the structure of AC in the system model is similar to the channel of two-receiver broadcast channels with confidential messages. When it comes to the discussion on broadcast channels, degraded, less noisy, and more capable channels are three important classes of channels that are often discussed because the single-letter characterization for the capacity region of these types of broadcast channels is determinable \cite[Chapter 5]{GK}.
The class of degraded channels can be further subdivided into two classes: the classes of physically and statistically degraded channels. It is known that the latter class is larger than the former.
%, that is, if a broadcast channel is physically degraded, then it is also statistically degraded, but the converse does not hold in general.
In this section, we will take a look into each characterization of the capacity regions for these important channel classes.
%The definitions of these classes of channels will be provided later.

%More precisely, we show that it is capable of characterizing the capacity regions of the authentication systems under the classes of degraded and less noisy ACs channels for discrete sources by a single auxiliary {random variables}. For more capable channels, we provide a remark on this class of channels at the end of this section.}
Prior to the presentation of our main results, the formal definitions of physically and stochastically degraded channels, less noisy, and more capable channels \cite{GK} are defined. In order not to confuse with AC of the authentication systems, we denote the conditional probability of the channel of two-receiver broadcast channels as $P_{BC|A}(b,c|a)$ for $(a,b,c) \in \mathcal{A}\times\mathcal{B}\times\mathcal{C}$, and $P_{B|A}(b|a)$ and $P_{C|A}(c|a)$ correspond to the conditional marginal distributions of the broadcast channels.

\begin{Definition}
(Physically degraded channel). $(\mathcal{A}, P_{C|A},\mathcal{C})$ is physically degraded {with respect to} $(\mathcal{A}, P_{B|A}, \mathcal{B})$ if $P_{BC|A}(b,c|a) = P_{B|A}(b|a)\cdot P_{C|B}(c|b)$ for some transition probabilities $P_{C|B}$.

%\vspace{2mm}
\noindent{(}Stochastically degraded channel). We say that $(\mathcal{A}, P_{C|A}, \mathcal{C})$ is stochastically degraded {with respect to} $(\mathcal{A}, P_{B|A},\mathcal{B})$ if there exists a channel $(\mathcal{B}, P_{C|B},\mathcal{C})$ such that ${P_{C|A}(c|a)} = \sum_{b \in \mathcal{B}}P_{C|B}(c|b)P_{B|A}(b|a)$.

%\vspace{2mm}
\noindent{(}Less noisy channel). $(\mathcal{A}, P_{B|A}, \mathcal{B})$ is less noisy than $(\mathcal{A},P_{C|A},\mathcal{C})$ if $I(B;{W}) \ge I(C;{W})$ for every {random variables} {$W$} such that ${W}-A-(B,C)$.

%\vspace{2mm}
\noindent{(}More capable channel). {$(\mathcal{A}, P_{B|A}, \mathcal{B})$ is more capable than $(\mathcal{A},P_{C|A},\mathcal{C})$ if $I(A;B) \ge I(A;C)$ for all $P_A$.}
\qed
\end{Definition}

{A clear relation among these classes of channels is that degraded channels are a subclass of less noisy channels, and less noisy channels are a subclass of more capable channels.}

In some literature, e.g., \cite{BB}, less noisy channels are called {noisier channels}. More precisely, it is said that $(\mathcal{A},P_{C|A},\mathcal{C})$ is noisier than $(\mathcal{A}, P_{B|A}, \mathcal{B})$ if for every {random variables} {$W$} such that ${W}-A-(B,C)$, we have that $I(B;{W}) \ge I(C;{W})$. In this manuscript, we sometimes use the terms ``less noisy channels'' and ``noisier channels'' interchangeably.

In order to simplify the statement of our main results, we define five new rate regions. The following rate constraints are used in the newly defined rate regions.
\begin{align}
R_S &\leq I(Y;U|Z), \label{rsdef3} \\
R_S &\leq I(Y;U)-I(Z;U), \label{rsdef32} \\
R_J &\geq I(\Tilde{X};U|Y), \label{rjdef3} \\
R_J &\geq I(\Tilde{X};U|Z), \label{rjdef32} \\
R_L &\geq I(X;U|Y) + I(X;Z). \label{rldef3}
\end{align}
\begin{Definition} \label{def333} {Rate} regions of secret-key, storage, and privacy-leakage rates {are} defined as
\vspace{-2mm}
\begin{align}
\mathcal{A}_1 = &\bigcup_{P_{U|\Tilde{X}}}\Big\{(R_S,R_J,R_L)\in \mathbb{R}^3_+:~{\rm The~ auxiliary~random} \nonumber \\
&{\rm variable}~U~{\rm satisfies~\eqref{rsdef3},~\eqref{rjdef3},~and~\eqref{rldef3}}\Big\}, 
\label{def31} \\
\mathcal{A}_2 = &\bigcup_{P_{U|\Tilde{X}}}\Big\{(R_S,R_J,R_L)\in \mathbb{R}^3_+:~{\rm The~ auxiliary~random} \nonumber \\
&{\rm variable}~U~{\rm satisfies~\eqref{rsdef3},~\eqref{rjdef32},~and~\eqref{rldef3}}\Big\}, \label{def32} \\
\mathcal{A}_3 = &\bigcup_{P_{U|\Tilde{X}}}\Big\{(R_S,R_J,R_L)\in \mathbb{R}^3_+:~{\rm The~ auxiliary~random} \nonumber \\
&{\rm variable}~U~{\rm satisfies~\eqref{rsdef32},~\eqref{rjdef3},~and~\eqref{rldef3}}\Big\}, \label{def33} \\
\mathcal{A}_4 = &\bigcup_{P_{U|\Tilde{X}}}\Big\{(R_S,R_J,R_L)\in \mathbb{R}^3_+:~{\rm The~ auxiliary~random} \nonumber \\
&{\rm variable}~U~{\rm satisfies~\eqref{rsdef32},~\eqref{rjdef32},~and~\eqref{rldef3}}\Big\},\label{def34}
\end{align}
where auxiliary random variable $U$ in the regions \eqref{def31} and \eqref{def32} satisfies the Markov {chain} $U-\Tilde{X}-X-Y-Z$ and auxiliary random variable $U$ in the regions \eqref{def33} and \eqref{def34} satisfies $U-\Tilde{X}-X-(Y,Z)$. The cardinality of the alphabet $\mathcal{U}$ on the auxiliary {random variables} $U$ in all regions above is constrained by $|\mathcal{U}| \le |\Tilde{\mathcal{X}}| + 3$. Also, define
\begin{align}
\mathcal{A}_5 = \{(R_S,R_J,R_L):~R_S = 0,~R_J \geq 0,~
R_L \geq I(X;Z)\}. \label{def5}
\end{align}
\qed
\end{Definition}

The regions $\mathcal{A}_1$ and $\mathcal{A}_2$ in Definition \ref{def333} correspond to the capacity regions of {the} GS and CS models when the {AC is} physically or statistically degraded, and the regions $\mathcal{A}_3$ and $\mathcal{A}_4$ are related to the capacity regions of {the} GS and CS models for less noisy ACs. The region $\mathcal{A}_5$ is used in {a} special case for degraded, less noisy, and Gaussian ACs, and no auxiliary {random variable} is involved in the expression in this region.

We start presenting our main results by showing a theorem when AC is degraded.

\begin{Theorem}\label{th1}
Suppose that AC $P_{YZ|X}$ has a structure such that Eve's channel $P_{Z|X}$ is physically degraded {with respect to}\ the main channel $P_{Y|X}$, meaning that the Markov chain $X-Y-Z$ holds. The {capacity regions} among secret-key, storage, and privacy-leakage rates of the GS and CS models are given by
\vspace{-2mm}
\begin{align}
\mathcal{R}_G = \mathcal{A}_1,~~~~~~\mathcal{R}_C = \mathcal{A}_2. \label{theorem1}
\end{align}
Reciprocally, if the Markov chain $X-Z-Y$ {holds}, the {regions are} characterized as
\vspace{-2mm}
\begin{align}
\mathcal{R}_G = \mathcal{R}_C = \mathcal{A}_5. \label{theorem2}
\end{align}
\qed
\end{Theorem}

The proof of Theorem \ref{th1} is similar to that of Theorem \ref{th2}; therefore, it is omitted.
\begin{Remark}
The capacity regions of physically and stochastically degraded ACs are given in the same form as in Theorem 1. This is because the capacity region depends on the marginal distributions $(P_{\Tilde{X}|X}, P_{Y|X}, P_{Z|X})$, and for the model considered in this paper, these distributions coincide for both physically and stochastically degraded ACs.
\end{Remark}

The following theorem states the capacity regions of the GS and CS models for less noisy ACs.
%which is the most important result in this paper.

\begin{Theorem}\label{th2}
If AC $P_{YZ|X}$ has a structure such that $P_{Y|X}$ is less noisy than $P_{Z|X}$, i.e., $I(Y;{W}) \ge I(Z;{W})$ for every {random variable} ${W}$ such that ${W}-X-(Y,Z)$, we have that
\begin{align}
\mathcal{R}_G = \mathcal{A}_3,~~~~~~\mathcal{R}_C = \mathcal{A}_4. \label{theorem3}
\end{align}
For the case where {$P_{Z|X}$ is less noisy than $P_{Y|X}$, i.e., $I(Y;{W}) \le I(Z;{W})$} for every ${W}$ such that ${W}-X-(Y,Z)$, the capacity {regions} of the systems {are} provided by
\begin{align}
\mathcal{R}_G = \mathcal{R}_C = {\mathcal{A}_5}. \label{theorem44}
\end{align}
\qed
\end{Theorem}

The proof of Theorem \ref{th2} is available in Appendix A. By a similar method used in \cite[Section V-A]{itw3}, it can be checked that both $\mathcal{R}_G$ and $\mathcal{R}_C$ are convex. In case of no presence of Eve ($Z$ is independent of other {random variables}), Theorems \ref{th1} and \ref{th2} naturally reduce to the {capacity regions} given in \cite{gunlu2018}.

Note that the assumption of less noisy channels seen in {Theorem \ref{th2}}, i.e., $I(Y;U) \ge I(Z;U)$ (or $I(Y;U) \le I(Z;U)$), is satisfied for every $U$ satisfying the Markov chain $U-\Tilde{X}-X-(Y,Z)$. This fact is utilized in the proof of this theorem.

\begin{Remark} \label{remarkmc}
The class of more capable channels includes less noisy channels as a special case \cite{GK}. When the {AC is} in the class of more capable {channels}, i.e., $I(X;Y) \ge I(X;Z)$ or $I(X;Y) \le I(X;Z)$, it is not yet known whether the capacity region can be characterized by one auxiliary random variable. More specifically, due to the impact of noise on {the} enrollment phase, the condition $I(X;Y) \ge I(X;Z)$ does not guarantee that $I(Y;U) \ge I(Z;U)$ and $I(\Tilde{X};Y) \ge I(\Tilde{X};Z)$, making it difficult to identify the sign in the right-hand side of the secret-key rate constraint in \eqref{rsdef32}. The same observation applies {to} the case in which $I(X;Y) \le I(X;Z)$.
\end{Remark}

An observation from the theorems and remarks shown above is that in the wiretap channels, the fundamental limits, e.g, the capacity-equivocation regions, depend on the channel $P_{YZ|X}$ only through the marginal distributions of the main channel $P_{Y|X}$ and Eve's channel $P_{Z|X}$ \cite{liang2009}. This conclusion may be applicable to a visible source model of the authentication systems. However, for the settings of {hidden source model}, the capacity regions are hinged on by not only the marginal distributions of AC {$P_{YZ|X}$} but also the EC {$P_{\Tilde{X}|X}$}.

\section{Examples} \label{sect4}
\subsection{Binary Sources}
In this section, the characterization of a binary example for Theorem \ref{th2} in the case where Eve's channel is noisier than the main channel is presented.

\begin{figure}[!t]
    \centering
    \includegraphics[scale=0.45]{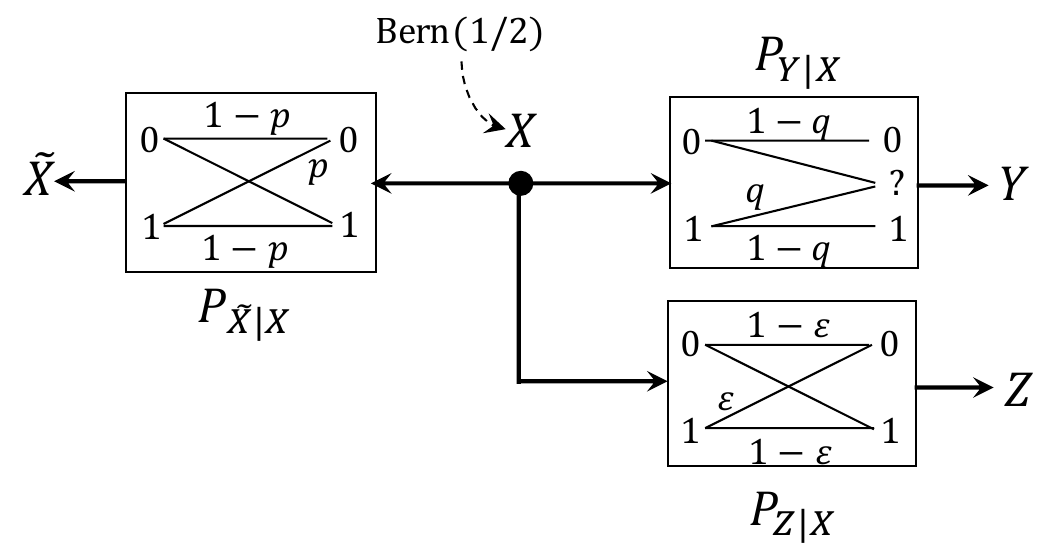}
    \vspace{-3mm}
    \caption{Transition probabilities of each channel for binary example}
    \label{binary}
    \vspace{-5mm}
\end{figure}

Consider the source {random variable} $X\sim \mathrm{Bern}(\frac{1}{2})$, $P_{\Tilde{X}|X}$ is a binary symmetric channel with crossover probability $p \in [0,1/2]$, $P_{Y|X}$ is a binary erasure channel with an erasure probability $q \in [0,1]$, and $P_{Z|X}$ is a {binary symmetric channel} with crossover probability
%\footnote{Different to the transition probability of the channel $P_{\Tilde{X}|X}$, where $p$ can be any value in $[0,1]$, the crossover probability $\epsilon$ can not surpass $\frac{1}{2}$. When it surpasses, Theorem \ref{th3} does not hold because the simple key lemma used for establishing this theorem is not provable.}
$\epsilon\in [0,1/2]$. Note that if the relation of $\epsilon$ and $q$ is such that $2q < \epsilon < 4q(1-q)$, $P_{Y|X}$ is less noisy than $P_{Z|X}$, but $P_{Y|X}$ is not a degraded version of $P_{Z|X}$. An illustration of this setting is described in {Figure} \ref{binary}.

Let the test channel $P_{U|\Tilde{X}}$ be a {binary symmetric channel} with crossover probability $\beta \in [0,1/2]$. The optimal rate regions of {the} GS and CS models in this case are given below.
\begin{Theorem} \label{th3} For binary sources when the main channel is less noisy than Eve's channel, the capacity regions of {the} GS and CS models are given as
\begin{align}
    \mathcal{R}_G =&\bigcup_{0 \le \beta \le 1/2}\Big\{(R_S,R_J,R_L)\in \mathbb{R}^3_+: \nonumber \\
    &R_S \le H_b(\beta*p*\epsilon) - (1-q)H_b(\beta*p) - q, \nonumber \\
    &R_J \ge q + (1-q)H_b(\beta*p) - H_b(\beta), \nonumber \\
    &R_L \ge 1+q -qH_b(\beta*p) - H_b(\epsilon)\Big\}, \label{bina-rg} \\
\mathcal{R}_C =&\bigcup_{0 \le \beta \le 1/2}\Big\{(R_S,R_J,R_L)\in \mathbb{R}^3_+: \nonumber \\
    &R_S \le H_b(\beta*p*\epsilon) - (1-q)H_b(\beta*p) - q, \nonumber \\
    &R_J \ge H_b(\beta*p*\epsilon) - H_b(\beta), \nonumber \\
    &R_L \ge 1+q -qH_b(\beta*p) - {H_b(\epsilon)} \Big\}, \label{bina-rc}
\end{align}
where the convolution operation $*$ is defined as $x*y = x(1-y) + (1-x)y$ for $x \in [0,1]$ and $y \in [0,1]$. \qed
\end{Theorem}
The proof of Theorem \ref{th3} is given in Appendix C. In \cite{gksc2018}, the rate region of {the} GS model for binary sources was derived under the assumptions that $\Tilde{X}=X$ (EC is noiseless) and the {AC is} physically degraded, i.e., the Markov chain $X-Y-Z$ holds. Theorem \ref{th3} is provided under a more general setting, and the key idea for deriving this theorem is to apply {Mrs. Gerber’s Lemma} \cite{wyner} in the reverse direction of Eve's channel to obtain an upper bound on the conditional entropy $H(Z|U)$. However, the obtained bound is not yet tight. We establish a simple lemma (Lemma \ref{binarylemma}) to acquire the optimal upper bound on $H(Z|U)$ to match the outer region with the inner region.

\subsection{Scalar Gaussian Sources} \label{gauss-source}
\begin{figure}[!t]
    \centering
    \vspace{-3mm}
    \includegraphics[scale=0.7]{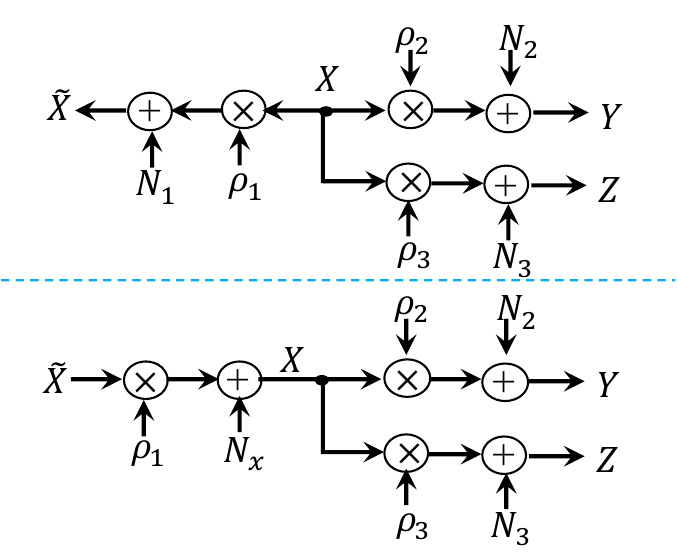}
    \vspace{-4mm}
    \caption{Data flows of the original system model (top) and transformed one (bottom) for Gaussian sources and channels}
    \label{guassian}
    \vspace{-5mm}
\end{figure}

Unlike the discrete sources, for Gaussian sources, we provide the capacity regions of the system for a general class of Gaussian ACs. A picture of data flows for Gaussian sources is depicted at the top of {Figure} \ref{guassian}. Assume that the source is given by $X \sim \mathcal{N}(0,1)$, and the channels $P_{\Tilde{X}|X}$, $P_{Y|X}$, and $P_{Z|X}$ are modeled as
\vspace{-1mm}
\begin{align}
    \Tilde{X} = \rho_1X + N_1,~Y = \rho_2X + {N_2},~Z = \rho_3X + {N_3}, \label{zyxn12}
\end{align}
where $|\rho_1|,|\rho_2|,|\rho_3| < 1$ are the correlation coefficients of each channel, $N_1 \sim \mathcal{N}(0,1-\rho^2_1)$, ${N_2} \sim \mathcal{N}(0,1-\rho^2_2)$, and ${N_3} \sim \mathcal{N}(0,1-\rho^2_3)$ are Gaussian {random variables}, and independent of each other and of other {random variables}.

Using a technique of transforming the exponent part of the joint distributions used in \cite{vy2} or covariance matrix transformations in \cite[Appnedix C.1]{wataoha2010}, \eqref{zyxn12} can be rewritten as
\vspace{-1mm}
\begin{align}
    X = \rho_1\Tilde{X} + N_x,~Y = \rho_2X + {N_2},~Z = \rho_3X + {N_3}, \label{zyxn123}
\end{align}
where $N_x \sim \mathcal{N}(0,1-\rho^2_1)$ and is independent of other {random variables}. A depiction of the data flows for \eqref{zyxn123} is displayed at the bottom of {Figure} \ref{guassian}, and the capacity regions for Gaussian sources are derived via \eqref{zyxn123} instead of \eqref{zyxn12}. The result is given below.

\begin{Theorem}\label{th4}
Under the condition of $\rho^2_2 > \rho^2_3$, i.e., $\Tilde{X}-X-Y-Z$ (cf.\ \cite[Lemma 6]{wataoha2010}), the capacity regions of the GS and CS models for Gaussian sources are given by
\vspace{-2mm}
\begin{align}
\mathcal{R}_G = &\bigcup_{P_{U|\Tilde{X}}}\Big\{(R_S,R_J,R_L)\in \mathbb{R}^3_+:~{\rm The~ auxiliary~random} \nonumber \\
&{\rm variable}~U~{\rm satisfies~\eqref{rsdef3},~\eqref{rjdef3},~and~\eqref{rldef3}}\Big\}, \label{gau123} \\
\mathcal{R}_C = &\bigcup_{P_{U|\Tilde{X}}}\Big\{(R_S,R_J,R_L)\in \mathbb{R}^3_+:~{\rm The~ auxiliary~random} \nonumber \\
&{\rm variable}~U~{\rm satisfies~\eqref{rsdef3},~\eqref{rjdef32},~and~\eqref{rldef3}}\Big\},
\label{gau1234}
\end{align}
where auxiliary random variable $U$ satisfies the Markov chain $U-\Tilde{X}-X-Y-Z$. Unlike Theorem \ref{th1}, the random variable $U$ is a continuous {random variable} and its cardinality is unbounded.
For the case of $\rho^2_2 \le \rho^2_3$, i.e., $\Tilde{X}-X-Z-Y$, the regions are characterized in the same form
\begin{align}
\mathcal{R}_G = \mathcal{R}_C = {\mathcal{A}_5}. \label{theorem233}
\end{align}
\qed
\end{Theorem}

Theorem \ref{th4} can be proved by a similar method {for} deriving Theorem \ref{th2}, and thus we omit the detailed proof. In {Theorems} \ref{th1}, \ref{th2}, and \ref{th4}, when the structure of ACs is {such} that the main channel is degraded {with respect to} Eve's channel or is noisier than Eve's one, the capacity regions of {the} GS and CS models are given in the same form. The secret-key generation at a positive rate is not possible, and the minimum value of the storage rates is zero, but that of the privacy-leakage rate can still be positive depending on the joint marginal densities of $(X,Z)$. Even when the encoding procedure is not needed, e.g., $U$ is set to be a constant, the information leaked to Eve via her channel $P_{Z|X}$ is at minimum rate $I(Z;X)$, which is equal to the capacity of this channel. This quantity corresponds to {an} uncontrollable amount of the privacy-leakage rate at the encoder, and it is avoidable if the privacy-leakage rate is {constrained} by conditional mutual information, {i.e., $I(X^n;J|Z^n)$}, as in \cite{gunlu2022}.

Note that due to the unbounded cardinality of the auxiliary {random variable}, the regions in \eqref{gau123} and \eqref{gau1234} are not directly computable. Next, we show that the parametric forms, i.e., computable expressions, of Theorem \ref{th4} are determined by a single parameter. The parameter $\alpha$ which appears in the following corollary acts as an adjusting parameter for the variance of the auxiliary {random variable} $U$. Unlike random variables $(\Tilde{X},X,Y,Z)$, in which {their} variances are always {one}, the auxiliary {random variable} $U$ could be any Gaussian random variable with a variance in the range $(0,1]$.

\begin{Corollary} \label{coroll1}
When the condition $\rho^2_2 > \rho^2_3$ is satisfied, we can compute the regions in {\eqref{gau123} and \eqref{gau1234}} as
\begin{align}
\mathcal{R}_G &= \bigcup_{\alpha \in (0,1]}\Big\{(R_S,R_J,R_L)\in \mathbb{R}^+_3:~ \nonumber \\
&R_S \leq \frac{1}{2}\log\left(\frac{\alpha\rho^2_1\rho^2_3 + 1 - \rho^2_1\rho^2_3}{\alpha\rho^2_1\rho^2_2 + 1 - \rho^2_1\rho^2_2}\right),\nonumber \\
&R_J \geq \frac{1}{2}\log \left(\frac{\alpha\rho^2_1\rho^2_2 + 1 - \rho^2_1\rho^2_2}{\alpha}\right),\nonumber \\
&R_L \geq \frac{1}{2}\log \left(\frac{\alpha\rho^2_1\rho^2_2 + 1 - \rho^2_1\rho^2_2}{(\alpha\rho^2_1 + 1 - \rho^2_1)(1 - \rho^2_3)}\right)\Big\}, \label{corollary11} \\
\mathcal{R}_C &= \bigcup_{\alpha \in (0,1]}\Big\{(R_S,R_J,R_L)\in \mathbb{R}^+_3:~ \nonumber \\
&R_S \leq \frac{1}{2}\log\left(\frac{\alpha\rho^2_1\rho^2_3 + 1 - \rho^2_1\rho^2_3}{\alpha\rho^2_1\rho^2_2 + 1 - \rho^2_1\rho^2_2}\right),\nonumber \\
&R_J \geq \frac{1}{2}\log \left(\frac{\alpha\rho^2_1\rho^2_3 + 1 - \rho^2_1\rho^2_3}{\alpha}\right),\nonumber \\
&R_L \geq \frac{1}{2}\log \left(\frac{\alpha\rho^2_1\rho^2_2 + 1 - \rho^2_1\rho^2_2}{(\alpha\rho^2_1 + 1 - \rho^2_1)(1 - \rho^2_3)}\right)\Big\}, \label{corollary12}
\end{align}
respectively, and that of \eqref{theorem233} is given as
\begin{align}
\mathcal{R}_G = \mathcal{R}_C &= \Big\{(R_S,R_J,R_L):~R_S = 0,~~R_J \ge 0, \nonumber \\
&R_L \geq \frac{1}{2}\log\left(\frac{1}{1-\rho^2_3}\right)\Big\}. \label{corollary13}
\end{align}
\qed
\end{Corollary}

The full proof of Corollary \ref{coroll1} is available in \cite[Appendix D]{vyo2022itw} and the convexity of these regions is verified in \cite[Appendix E]{vyo2022itw}. When EC is noiseless ({i.e.,} $\rho^2_1 \rightarrow 1$), Corollary \ref{coroll1} reduces to the {parametric forms} derived in \cite[Corollary 1]{vyo2022}. In addition, when Eve can observe only the helper data, corresponding to the case in which $Z$ is independent of other {random variables} ($\rho^2_3 = 0$), Corollary \ref{coroll1} matches with the parametric expressions of {the} GS and CS models provided in \cite[Corollary 1]{vy2}.

\subsection{Behaviors of the Capacity Region for Gaussian Sources}
In this section, we investigate the ultimate (asymptotic) limits of the secret-key and privacy-leakage rates and provide some numerical results under Gaussian sources. For brevity, we focus only on the GS model.
%\footnote{\revb{Based on \eqref{gunlu333}, an achievable region (non-tight bound) involving two parameters with range $(0,1]$ is derived. Suppose that the range is equally sliced with step side $10^{-3}$. The total steps of calculating the region with two parameters are $(1/10^{-3}-1)(1/10^{-3}-1) = 998001$ steps, while that of \eqref{corollary11} is $(1/10^{-3}-1) = 999$, resulting in lower complexity. Also, the non-tight bound does not guarantee that the calculated tuple $(R_S,R_L,R_J)$ is optimal. Thus, in general, the region with two parameters is not used when calculating the theoretical behaviors of the capacity region.}}.
%with $\rho^2_2 > \rho^2_3$.

First, we find expressions for the {optimal} secret-key and privacy-leakage rates under a fixed condition of $R_J$ for the {hidden source model}. Let us fix the storage rate
\begin{align}
    R^{\alpha}_J = \frac{1}{2}\log \left(\frac{\alpha\rho^2_1\rho^2_2 + 1 - \rho^2_1\rho^2_2}{\alpha}\right), \label{rj-alpha}
\end{align}
equivalent to $\alpha = \frac{1-\rho^2_1\rho^2_2}{e^{2R^{\alpha}_J}-\rho^2_1\rho^2_2}$.
%When $\alpha \downarrow 0$, it is obvious that $R^{\alpha}_J \rightarrow \infty$.
Now define two rate functions
\begin{align}
    R^*_S(R^{\alpha}_J) = \max_{(R_S,R^{\alpha}_J,R_L) \in \mathcal{R}_G}R_S, \nonumber \\
    R^*_L(R^{\alpha}_J) = \min_{(R_S,R^{\alpha}_J,R_L) \in \mathcal{R}_G}R_L. \label{3535}
\end{align}
Using the value of $\alpha$, we can write that
\begin{align}
    R^*_S(R^{\alpha}_J) &= \frac{1}{2}\log\left(\frac{1-\rho^2_1\rho^2_3 - \rho^2_1(\rho^2_2-\rho^2_3)e^{-{2R^{\alpha}_J}}}{1-\rho^2_1\rho^2_2}\right),\nonumber \\
    R^*_L(R^{\alpha}_J) &= \frac{1}{2}\log\left(\frac{1-\rho^2_1\rho^2_2}{(1-\rho^2_3)(1-\rho^2_1 + \rho^2_1(1-\rho^2_2)e^{-2{R^{\alpha}_J}})}\right). \label{rsrlrj-star}
\end{align}
The asymptotic limits of secret-key and privacy-leakage rates when $R^{\alpha}_J$ {tends} to infinity are given by
\begin{align}
    \lim_{R^{\alpha}_J \rightarrow \infty}R^*_S(R^{\alpha}_J) &= \frac{1}{2}\log\left(\frac{1-\rho^2_1\rho^2_3}{1-\rho^2_1\rho^2_2}\right) = I(Y;\Tilde{X}|Z), \nonumber \\
    \lim_{R^{\alpha}_J \rightarrow \infty}R^*_L(R^{\alpha}_J) &= \frac{1}{2}\log\left(\frac{1-\rho^2_1\rho^2_2}{(1-\rho^2_1)(1-\rho^2_3)}\right) \nonumber \\
    &= I(X;\Tilde{X}|Y) + I(X;Z). \label{rsasym}
\end{align}

For the visible source model, the asymptotic limits of secret-key and privacy-leakage rates for a given storage rate are determinable by substituting $\rho^2_1 = 1$ into \eqref{rsasym}, i.e.,
\begin{align}
    \lim_{\Tilde{R}^{\alpha}_J \rightarrow \infty}\Tilde{R}^*_S(\Tilde{R}^{\alpha}_J) &= \frac{1}{2}\log\left(\frac{1-\rho^2_3}{1-\rho^2_2}\right) = I(X;Y|Z), \nonumber \\
    \lim_{\Tilde{R}^{\alpha}_J \rightarrow \infty}\Tilde{R}^*_L(\Tilde{R}^{\alpha}_J) &= \lim_{\Tilde{R}^{\alpha}_J \rightarrow \infty}\left(\frac{1}{2}\log\left(\frac{1}{1-\rho^2_3}\right) + \Tilde{R}^{\alpha}_J\right) \rightarrow \infty,
    \label{414141}
\end{align}
where $\Tilde{R}^{\alpha}_J = \frac{1}{2}\log \left(\frac{\alpha\rho^2_2 + 1 - \rho^2_2}{\alpha}\right)$ and $\Tilde{R}^*_S(\Tilde{R}^{\alpha}_J)$ and $\Tilde{R}^*_L(\Tilde{R}^{\alpha}_J)$ are defined in {the same} manner as \eqref{3535} and correspond to the maximum secret-key rate and the minimum privacy-leakage rate for this model under fixed $\Tilde{R}^{\alpha}_J$. One can see that in the second equation of \eqref{414141}, the optimal value of the privacy-leakage rate increases linearly with the storage rate.

\begin{comment}
Next, we look into special points of the {visible source model}. See \cite[Corollary 1]{vyo2022} for detailed characterizations. Similar to the above arguments, fix $\Tilde{R}^{\alpha}_J = \frac{1}{2}\log \left(\frac{\alpha\rho^2_2 + 1 - \rho^2_2}{\alpha}\right)$ for the {visible source model} and thus,
\vspace{-3mm}
\begin{align}
    \Tilde{R}^*_S(\Tilde{R}^{\alpha}_J) &= \frac{1}{2}\log\left(\frac{1-\rho^2_3 - (\rho^2_2-\rho^2_3)e^{-{2}\Tilde{R}^{\alpha}_J}}{1-\rho^2_2}\right), \nonumber \\
    \Tilde{R}^*_L(\Tilde{R}^{\alpha}_J) &= \frac{1}{2}\log\left(\frac{1}{1-\rho^2_3}\right) + \Tilde{R}^{\alpha}_J, \label{39eq}
\end{align}
where $\Tilde{R}^*_S(\Tilde{R}^{\alpha}_J)$ and $\Tilde{R}^*_L(\Tilde{R}^{\alpha}_J)$ are defined in {the same} manner of \eqref{3535} and correspond to the maximum secret-key rate and the minimum privacy-leakage rate for the {visible source model} under fixed $\Tilde{R}^{\alpha}_J$. Also, we can see that in the second equation of \eqref{39eq}, the optimal values of the privacy-leakage and storage rates are in a linear relation. The asymptotic limits of these rates as $\Tilde{R}^{\alpha}_J \rightarrow \infty$ are
\end{comment}

Next, we provide some numerical calculations of the region $\mathcal{R}_G$ in \eqref{corollary11}, and take a look into special points of both the visible source model ($\rho^2_1 = 1$) and hidden source model $(\rho^2_1 < 1)$. The following two scenarios are considered.
\begin{itemize}
    \item[1)] $\rho^2_1$ varies over three values $1.0$, $0.9$, and $0.7$, but $(\rho^2_2,\rho^2_3)$ is fixed at $(0.8,0.5)$. This is the case where the probability of enrollment channel $P_{\Tilde{X}|X}$ could be changed, but that of the authentication channel $P_{YZ|X}$ remains the same.
    \item[2)] $\rho^2_1$ is fixed at $0.9$, but $(\rho^2_2,\rho^2_3)$ could be either one of the pairs $(0.8,0.5)$, \revb{$(0.7,0.6)$, or $(0.6,0.7)$}. This is the opposite example of Scenario 1).
\end{itemize}
%calculate four cases in which Case 1) $\rho^2_1 = 1$, Case 2) $\rho^2_1 = 0.9$, Case 3) $\rho^2_1 = 0.7$, and Case 4) $\rho^2_1 = 0.5$. We fix $\rho^2_2 = 0.7$ and $\rho^2_3=0.3$ in all cases.
%Note that $\rho^2_2 > \rho^2_3$ is satisfied in each scenario.
%Literally, Scenarios 1) and 2) are contrastive examples where the correlation coefficient and noise of EC are changed in Scenario 1) but these elements are fixed in Scenario 2).

\begin{figure*}[!t]
\centering
\begin{minipage}[!t]{.47\textwidth}
\includegraphics[scale=0.6]{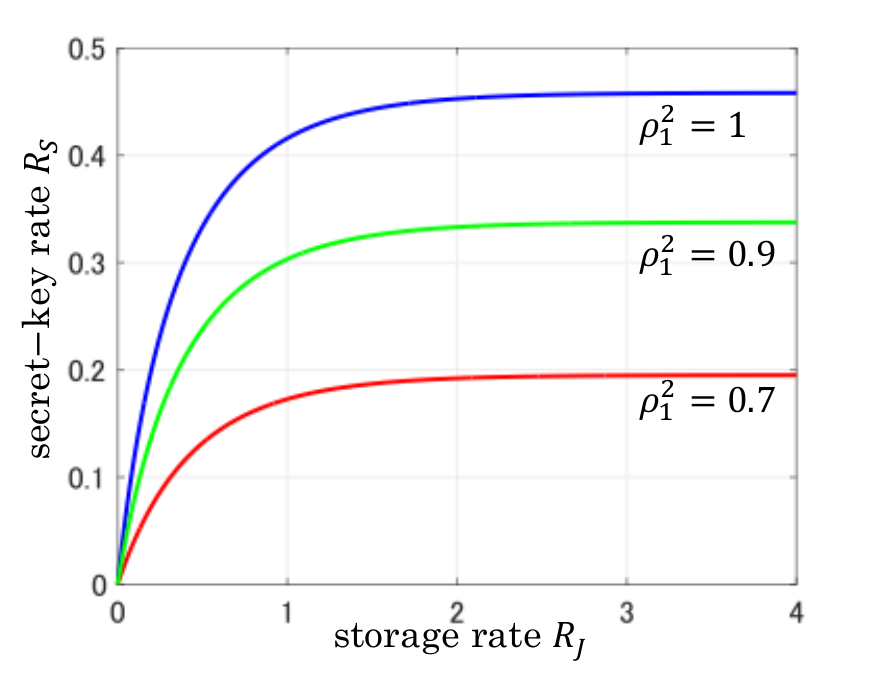}
\vspace{-7mm}
\caption{Projection of the capacity region $\mathcal{R}_G$ in \eqref{corollary11} with different $\rho^2_1$ onto $R_JR_S$-plane.}
\label{label-a}
\end{minipage}\quad
\begin{minipage}[!t]{.47\textwidth}
\includegraphics[scale=0.6]{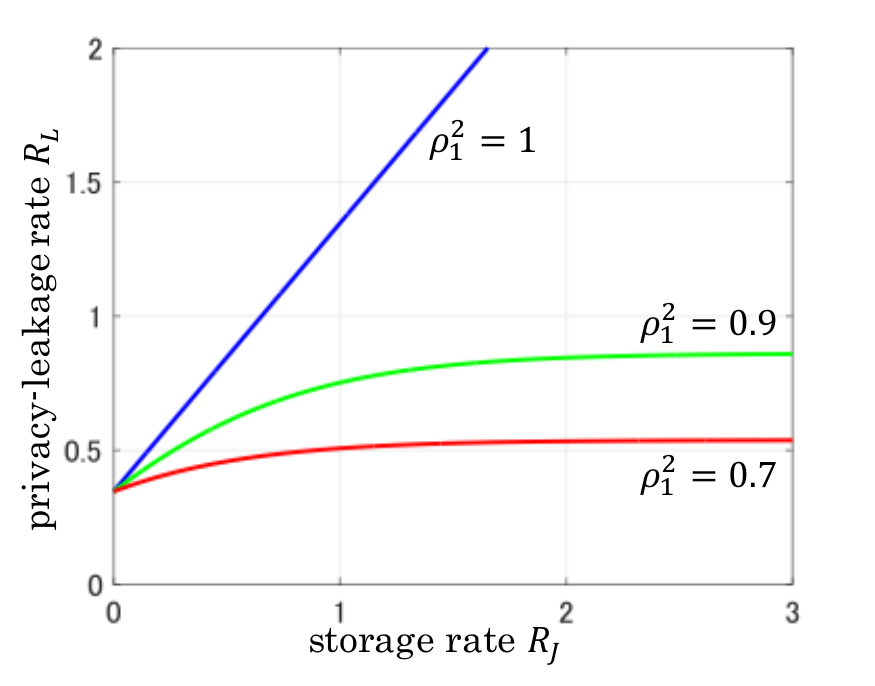}
\vspace{-7mm}
\caption{Projection of the capacity region {$\mathcal{R}_G$} in \eqref{corollary11} with different $\rho^2_1$ onto $R_JR_L$-plane.}
\label{label-b}
\end{minipage}
\begin{minipage}[!t]{.47\textwidth}
\includegraphics[scale=0.6]{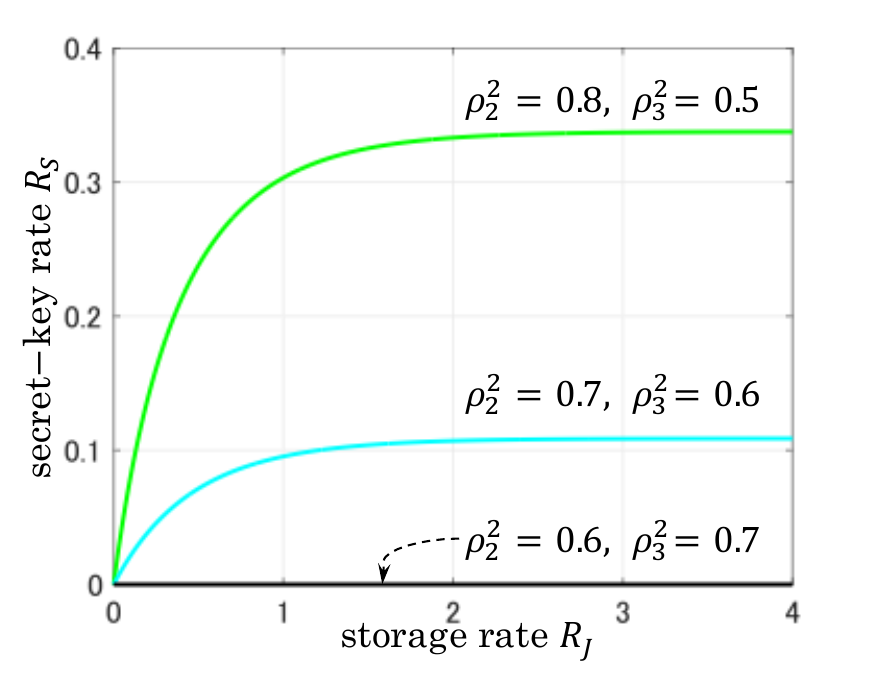}
\vspace{-7mm}
\caption{\rev{Projection of the capacity region $\mathcal{R}_G$ in \eqref{corollary11} with different $\rho^2_2$ and $\rho^2_3$ onto $R_JR_S$-plane.}}
\label{label31}
\end{minipage}\quad
\begin{minipage}[!t]{.47\textwidth}
\includegraphics[scale=0.6]{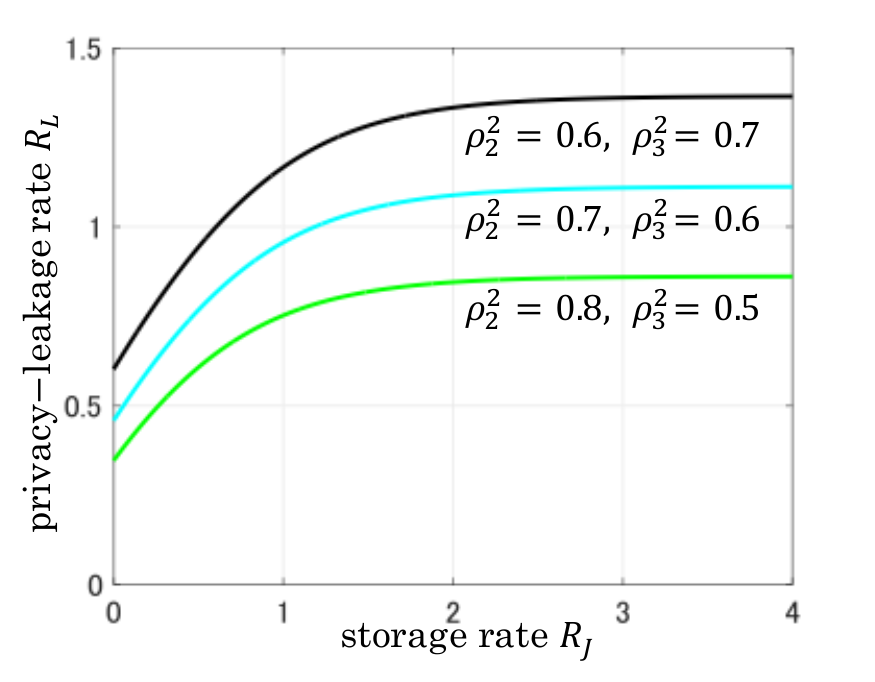}
\vspace{-7mm}
\caption{\rev{Projection of the capacity region $\mathcal{R}_G$ in \eqref{corollary11} with different $\rho^2_2$ and $\rho^2_3$ onto $R_JR_L$-plane.}}
\label{label32}
\end{minipage}
\vspace{-5mm}
\end{figure*}

\revb{Figures \ref{label-a} and \ref{label-b} plot the optimal values between secret-key versus storage rates $(R^{\alpha}_J,R^*_S(R^{\alpha}_J))$ and privacy-leakage versus storage rates $(R^{\alpha}_J,R^*_L(R^{\alpha}_J))$, respectively, for Scenario 1). Figures \ref{label31} and \ref{label32} illustrate the relations of the same rate pairs for Scenario 2). These figures are obtained by calculating the values of $R^{\alpha}_J$ defined in \eqref{rj-alpha}, and $R^*_S(R^{\alpha}_J)$ and $R^*_L(R^{\alpha}_J)$ defined in \eqref{rsrlrj-star} with respect to the parameter $\alpha$. In this calculation, we set the step size of $\alpha$ to be $10^{-5}$, which was found to be sufficiently small for numerical implementation}.

From Figures \ref{label-a} and \ref{label-b}, the visible source model produces a better secret-key rate but leaks more privacy of physical identifiers to Eve compared to the performances of the hidden source model. More precisely,
%In these figures, the blue, red, green, and black curves show the theoretical behaviors regarding the {hidden source model} when $\rho^2_1 = \frac{7}{8}$ and $\rho^2_1 = \frac{31}{32}$, respectively, and the blue curve illustrates the calculated result of the visible source model. Overall,
%In both Cases 1) and 2), the graphs of secret-key versus storage rates and privacy-leakage versus storage rates for the {visible source model} are duplicated since the correlation coefficient $\rho^2_1$ does not engage in these asymptotic values (cf. \eqref{414141}) and the values of $\rho^2_2, \rho^2_3$ in both cases are the same. However, this is not applied to the {hidden source model}.
the asymptotic values of secret-key rate are $\lim_{R^{\alpha}_J \rightarrow \infty}R^*_S(R^{\alpha}_J) = \frac{1}{2}\log (\frac{5}{2}) = 0.458$ nats, $0.338$ nats, and $0.195$ nats when $\rho^2_1$ is equal to $1.0$, $0.9$, and $0.7$, respectively. This indicates that when $\rho^2_1$ decreases, implying that the noise introduced to the identifiers in the enrollment phase increases, the secret-key rate becomes smaller.

Conversely, in terms of the privacy-leakage rate, the asymptotic limits become $\lim_{R^{\alpha}_J \rightarrow \infty}R^*_L(R^{\alpha}_J) \rightarrow \infty$, $0.861$ nats, and $0.538$ nats when $\rho^2_1$ is equal to $1.0$, $0.9$, and $0.7$, respectively. Evidently, when $\rho^2_1$ is low, less information about the identifiers leaks to Eve. By the reason that the noise in the enrollment phase serves as a fixed filter \cite{zhou2021} to obscure the privacy of identifiers, when $\rho^2_1$ is small (the variance of noise added to the identifiers in the EC is large), the amount of information leaked to Eve is also small. By contrast, when $\rho^2_1$ approaches $1$, the effectiveness of the filter is lessened, and the {hidden source model} behaves similarly to the visible source model. Thus, a larger amount of privacy of the identifiers could be leaked.

For Scenario 2), \revb{Figures \ref{label31} and \ref{label32} show that the achievable secret-key rate gradually decreases and the privacy-leakage rate rises as the value of $\rho^2_2$ declines and that of $\rho^2_3$ increases, which can explicitly be verified by comparing two distinct values of the secret-key and privacy-leakage rates in \eqref{rsrlrj-star} with different pairs $(\rho^2_2,\rho^2_3)$ under the same storage rate}. These behaviors suggest that when the noise variance of the measurements observed through the main channel is large, corresponding to the case where a low-quality quantizer, e.g., quantizer with few quantization levels, is deployed at the decoder, it leads to a small secret-key generation rate and a high privacy-leakage rate. This effect becomes particularly remarkable when Eve uses a high-quality quantizer.

In the authentication systems, it is favored for achieving a high secret-key rate while maintaining low storage and privacy-leakage rates, but these simulation results reflect the difficulty of achieving such a tuple simultaneously. Therefore, to prevent a circumstance such that a significant loss of privacy occurs, it may be important not only to focus on increasing the gain for the secret-key rate but also to weigh its balance with the storage and privacy-leakage rates as well when designing practical codes for an authentication system.
%Last, from Figures \ref{label-a}--\ref{label32}, it is evident that \rev{the noises in the enrollment and authentication phases} have a direct impact on the achievable region of the authentication systems.

\section{Conclusion} \label{sect5}
In this paper, we investigated the classes of ACs {for which} the capacity regions of the {GS and CS models} can be characterized by one auxiliary random variable. The obtained results revealed that only a single auxiliary random variable is required to characterize the capacity regions for degraded and less noisy ACs. Moreover, the capacity regions of the authentication systems for both binary and Gaussian sources were derived. \revb{All the expressions derived in this work are not only tight but also readily computable. They may serve as a performance benchmark when practical channel codes such as LDPC and polar codes are constructed for the authentication systems as in \cite{chen2019} for a visible source model.} We also provided some numerical calculations for the Gaussian case to demonstrate the impact of noise in the enrollment phase on the capacity region as well as to examine the trade-off between secret-key and privacy-leakage rates for a given storage rate.

\revb{For future work, a natural extension of this work is to investigate whether polar codes can achieve all the rate points in the capacity region of binary sources. In fact, for the typical key-agreement problem \cite{chou2015}, polar codes were shown to achieve the fundamental limits by exploiting the degraded and less noisy properties of the main and Eve's channels. Due to the similarities of the key-agreement problem and our model, it may be possible to demonstrate that the code achieves the fundamental limits of the authentication systems as well}. Furthermore, extending the results in Section \ref{gauss-source} to vector Gaussian case is another interesting research topic.
%Moreover, considering two-encoder source coding problem for the authentication systems is of sufficient interest.
%In this setting, not only $\Tilde{X}^n$ but also $Y^n$ are encoded in order to minimize the communication rate in the authentication phase.
%In \cite{Naghibi2015}, an HSM without estimating the secret key, and the privacy constraint is measured by the conditional entropy of the source sequence $X^n$ given the helper data $J$ and the sequence $Z^n$ was analyzed. As an extension of the model in \cite{Naghibi2015}, it may still be of interest to clarify the capacity region of the systems under the setting when the privacy constraint is imposed by the mutual information as in this paper.

\appendices
\section{Proof of Theorem \ref{th2}}
This appendix deals with the proof of the capacity regions for less noisy ACs. We only provide the proof of \eqref{theorem3} since that of \eqref{theorem44} follows similarly by simply setting the auxiliary {random variable} $U$ to be constant. The entire proof is divided into two parts, namely, the converse and achievability parts. For the converse part, the derivation of each rate constraint for the GS and CS models is discussed in detail, while in the achievability part, only the key point is addressed.

\subsection{Converse Part}
\begin{figure}[t]
\centering
\includegraphics[scale=0.55]{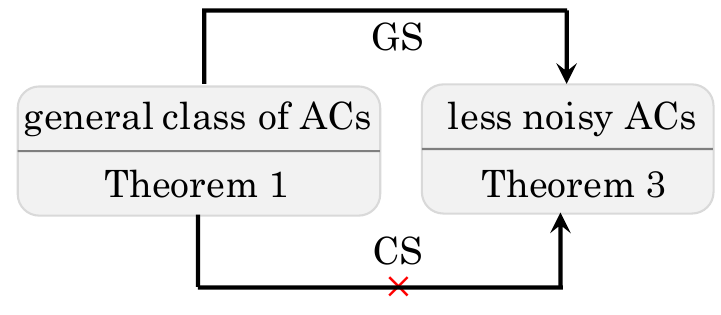}
\vspace{-2mm}
\caption{The possibility of a reduction in Theorem \ref{gunlu2018} to obtain the outer regions of GS and CS models less noisy ACs.}
\label{c1}
\end{figure}

\begin{figure}[!t]
    \centering
\includegraphics[scale=0.54]{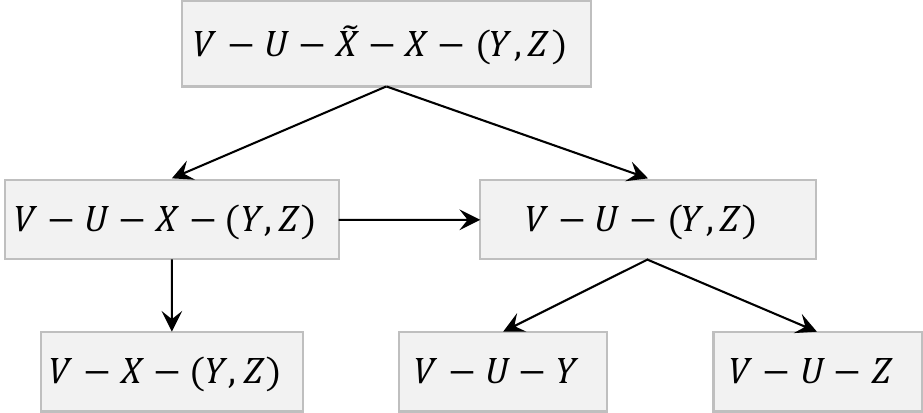}
\caption{{Some shorter Markov chains that can be derived from the long Markov chain in Theorem \ref{gunlu2018}}.}
\label{c2}
\vspace{-3mm}
\end{figure}

Note that following the same technique used in \cite{gksc2018}, the capacity regions derived under the general class of ACs also {hold} for less noisy ACs. {Figure \ref{c1} illustrates the possibility of a direct deduction of the capacity regions of the GS and CS models for less noisy ACs via the expressions with two auxiliary {random variables} that we have seen in Theorem \ref{gunlu2018}.

More specifically, it is possible to derive the outer region on $\mathcal{R}_G$ in Theorem \ref{th2} directly via the region in \eqref{gunlu333} by exploiting the long Markov chain $V-U-\Tilde{X}-X-(Y,Z)$, as shown in Figure \ref{c2}, and the property of less noisy channels, but the same approach cannot be applied to the CS model.} In the proof, we demonstrate the proofs of {the} GS and CS models {via} different approaches. The proof begins with {the} GS model and follows by the detailed argument of {the} CS model.

\medskip
\noindent{\em  Converse Proof of GS Model}:~~~~Since the bounds {on} $R_J$ in both the regions in \eqref{gunlu333} and \eqref{theorem3} remain unchanged, we need to check the constraints on the secret-key and privacy-leakage rates. %Note that with an random variable $W$ such that $W-X-(Y,Z)$, we have that $I(Y;W) \ge I(Z;W)$ for less noisy ACs.
Transform the bound on the secret-key rate as follows:
\begin{align}
    R_S & \le I(Y;U|V) - I(Z;U|V) \nonumber \\
    &\overset{\mathrm{(a)}}= I(Y;U) - I(Y;V) - (I(Z;U)-I(Z;V)) \nonumber \\
    &= I(Y;U) - I(Z;U) - (I(Y;V)-I(Z;V)) \nonumber \\
    &\overset{\mathrm{(b)}}\le I(Y;U)-I(Z;U), \label{rsconverse}
\end{align}
where (a) follows by the Markov chains $V-U-Y$ and $V-U-Z$, {derivable from the Markov chain $V-U-\Tilde{X}-X-(Y,Z)$ (cf.\ Figure \ref{c2})}, and (b) follows because less noisy ACs fulfill the condition $I(Y;V) \ge I(Z;V)$ for every $V-X-(Y,Z)$.

Likewise, for the bound on {the} privacy-leakage rate, we have
\begin{align}
    R_L &\geq I(X;U,Y) - I(X;Y|V) + I(X;Z|V) \nonumber \\
        &\overset{\mathrm{(a)}}= I(X;U,Y) - {\left(I(X;Y) - I(Y;V)\right)}  \nonumber \\
        &~~~ + {I(X;Z)} - I(Z;V) \nonumber \\
        &= I(X;U|Y) + I(X;Y) - {\left(I(X;Y) - I(Y;V)\right)}  \nonumber \\
        &~~~ + {I(X;Z)} - I(Z;V) \nonumber \\
        &= I(X;U|Y) + {I(X;Z)} + I(Y;V) - I(Z;V) \nonumber \\
        &\overset{\mathrm{(b)}}\ge I(X;U|Y) + {I(X;Z)}, \label{rlconverse}
\end{align}
where (a) is due to the Markov chain $V-X-(Y,Z)$ and (b) is due to the property that $I(Y;V) \ge I(Z;V)$ for less noisy ACs. Hence, the converse proof of {the} GS model is attained.
\qed

\medskip
\noindent{\em Converse Proof of CS Model}:~~~~Observe that the right-hand side of the storage rate of {the} CS model with two auxiliary {random variables} can be reshaped as
\begin{align}
    R_J &\ge I(\Tilde{X};U|Y) + I(Y;U|V)-I(Z;U|V) \nonumber \\
    &\overset{\mathrm{(a)}}= I(\Tilde{X};U) - I(Y;U) + I(Y;U) - I(Y;V) \nonumber \\
    &~~~- (I(Z;U) - I({Z};V)) \nonumber \\
    &= I(\Tilde{X};U)-I(Z;U)- (I(Y;V) - I(Z;V)) \nonumber \\
    &\overset{\mathrm{(b)}}= I(\Tilde{X};U|Z) - (I(Y;V) - I(Z;V)), \label{rjconverse}
\end{align}
where (a) is due to the Markov chains $U-\Tilde{X}-Y$ and $V-U-(Y,Z)$, and (b) follows from the Markov chain $U-\Tilde{X}-Z$.

In \eqref{rjconverse}, since $I(Y;V) \ge I(Z;V)$ {for less noisy ACs}, this lower bound {cannot} be further reduced to the one seen in \eqref{rjdef32}. We cannot apply the technique used for {the} GS model to derive the outer bound directly from the region with two auxiliary {random variables} {(cf. {eq}. \eqref{gunlu333cs})} for the CS model, and thus an alternative approach is required. Here, we make use of a standard technique that relies on the assumption of auxiliary {random variables} and Fano's inequality.
%to analyze each rate constraint.

Suppose that a rate tuple $(R_S,R_J,R_L)$ is achievable, implying that there exists a pair of encoders and decoders such that all requirements in Definition 2 are satisfied {for small enough $\delta > 0$ and block length $n \ge n_0~(n_0 \ge 1)$}. {For $t \in [1:n]$}, define auxiliary {random variables} $U_t = (J,S,Y^n_{t+1},Z^{t-1})$ and $V_t = (J,Y^n_{t+1},Z^{t-1})$. {Under these settings, it is easy to verify} that the Markov chain $V_t-U_t-\Tilde{X}_t-X_t-(Y_t,Z_t)$ is satisfied.
%Assume that the rate tuple $(R_S,R_J,R_L)$ is achievable for small enough $\delta > 0$ and block length $n \ge n_0,~(n_0 \ge 1)$.
%$\delta_n = \frac{1}{n}(1+\delta\log M_S)$, and $\delta_n$ goes to zero as $\delta \downarrow 0$ and $n \rightarrow \infty$.

\medskip
\noindent{\em {Analysis of} Secret-key Rate}:~~~~Define $\delta_n = \frac{1}{n}(1+\delta\log M_S)$, and this quantity is related to an upper bound of Fano's inequality. From \eqref{cs-keycon}, the secret-key rate can be bounded by
\begin{align}
    n&(R_S - {\delta}) \le {\log M_S} =  H(S) \nonumber \\
    &\overset{\mathrm{(a)}}\le \sum_{t=1}^n\{I(Y_t;U_t|V_t) - I(Z_t;U_t|V_t)\} + n(\delta_n + \delta) \nonumber \\
    &\overset{\mathrm{(b)}}=\sum_{t=1}^n\{I(Y_t;U_t) - I(Z_t;U_t)-(I(Y_t;V_t) - I(Z_t;V_t))\} \nonumber \\
    &~~~~~~~~+ n(\delta_n + \delta) \nonumber \\
    &\overset{\mathrm{(c)}}\le \sum_{t=1}^n\{I(Y_t;U_t) - I(Z_t;U_t)\} + n(\delta_n + \delta), \label{nrs}
\end{align}
where (a) is due to a similar argument in \cite[{eq}. (40)]{gunlu2018}, (b) holds since the Markov chains $V_t-U_t-Y_t$ and $V_t-U_t-Z_t$ are applied, and (c) follows from the Markov {chain} $V_t-U_t-(Y_t,Z_t)$ and the property of less noisy channels that $I(Y_t;V_t) \ge I(Z_t;V_t)$ for any {random variable} $V_t$ such that $V_t-X_t-(Y_t,Z_t)$.

\medskip
\noindent{\em Analysis of Storage Rate}:~~~~For the CS model, note that the secret key $S$ is independent of random variables $(\Tilde{X}^n,X^n,Y^n,Z^n)$, and the helper data {$J$ is a function of $(\Tilde{X}^n,S)$}. {From \eqref{storage}, we} have that
\begin{align}
    &n(R_J+\delta) \ge \log M_J \ge H(J)
    = I(\Tilde{X}^n,S;J) \nonumber \\
    &= I(\Tilde{X}^n,S;J,Z^n) - I(\Tilde{X}^n,S;Z^n|J)\nonumber \\
    %&= I(S;J,Z^n) + I(\Tilde{X}^n;J,Z^n|S) \nonumber \\
    %&~~~- I(S;Z^n|J) - I(\Tilde{X}^n;Z^n|J,S)\nonumber \\
    &\ge I(\Tilde{X}^n;J,Z^n|S) - I(\Tilde{X}^n;Z^n|J,S) \nonumber \\
    &\overset{\mathrm{(a)}}= I(\Tilde{X}^n;J,Z^n|S) - H(Z^n|J,S) + H(Z^n|\Tilde{X}^n) \nonumber \\
    &\ge I(\Tilde{X}^n;J,Z^n|S) - I(\Tilde{X}^n;Z^n) \nonumber \\
    &= I(\Tilde{X}^n;J|Z^n,S) + I(\Tilde{X}^n;Z^n|S) - I(\Tilde{X}^n;Z^n) \nonumber \\
    &= I(\Tilde{X}^n;J|S,Z^n) = H(\Tilde{X}^n|Z^n) - H(\Tilde{X}^n|J,S,Z^n) \label{middleeq} \\
    &= \sum_{t=1}^n\{H(\Tilde{X}_t|Z_t)-H(\Tilde{X}_t|\Tilde{X}^{t-1},J,S,Z^n)\} \nonumber \\
    &\overset{\mathrm{(b)}}= \sum_{t=1}^n\{H(\Tilde{X}_t|Z_t)-H(\Tilde{X}_t|\Tilde{X}^{t-1},J,S,Z^n,Y^n_{t+1})\} \nonumber \\
    &\overset{\mathrm{(c)}}\ge \sum_{t=1}^n\{H(\Tilde{X}_t|Z_t)-H(\Tilde{X}_t|J,S,Y^n_{t+1},{Z^t})\} \nonumber \\
    &= \sum_{t=1}^n\{H(\Tilde{X}_t|Z_t)-H(\Tilde{X}_t|Z_t,U_t)\} = \sum_{t=1}^nI(\Tilde{X}_t;U_t|Z_t), \label{nrj222}
\end{align}
where (a) is due to the Markov chain $Z^n-(\Tilde{X}^n,S)-J$ and $S$ is independent of other {random variables}, (b) is due to the Markov chain $\Tilde{X}_t-(\Tilde{X}^{t-1},J,S,Z^n)-Y^n_{t+1}$, and (c) {follows because conditioning reduces entropy}.

\medskip
\noindent{\em {Analysis} of Privacy-Leakage Rate}:~~~~We can {develop} the right-hand side of \eqref{privacy} as
\begin{align}
    n(R_L &+ \delta) \ge I(X^n;J,Z^n) \nonumber \\
    &= I(X^n;J,S,Z^n) - I(X^n;S|J,Z^n) \nonumber \\
    &\ge I(X^n;J,S,Z^n) - H(S) \nonumber \\
    &= I(X^n;J,S|Z^n) + nI(X;Z) - H(S) \nonumber \\
    &= I(X^n;J|S,Z^n) + nI(X;Z) - H(S) \nonumber \\
    &\overset{\mathrm{(a)}}\ge \sum_{t=1}^nI(X_t;U_t|Z_t) + nI(X;Z) - H(S)\nonumber \\
    &\overset{\mathrm{(b)}}\ge \sum_{t=1}^nI(X_t;U_t|Y_t) + nI(X;Z) -n(\delta_n + \delta), \label{nrl}
\end{align}
where (a) follows from similar steps between \eqref{middleeq} and \eqref{nrj222}, and (b) {follows since $H(S)$ is upper bounded by} the last inequality in \eqref{nrs}.

The proof wraps up with the standard argument for single letterization using a time-sharing {random variable} $Q$, where the {random variable $Q$ is uniformly distributed on $[1:n]$ and} independent of other {random variables}. More specifically, define $X = X_Q$, $\Tilde{X} = \Tilde{X}_Q$, $Y = Y_Q$, $Z = Z_Q$, and $U = (U_Q,Q)$, so that $U-\Tilde{X}-X-(Y,Z)$ forms a Markov chain, and finally, letting {$n \rightarrow \infty$ and $\delta \downarrow 0$}, from \eqref{nrs}, \eqref{nrj222}, and \eqref{nrl}, {we obtain $\mathcal{R}_C \subseteq \mathcal{A}_4$}.

For the cardinality bound on the set $\mathcal{U}$ of the auxiliary random variable $U$, we apply the support lemma \cite[Lemma 3.4]{GK} to show that $|\mathcal{U}| \le |\Tilde{\mathcal{X}}| + 3$. More precisely, $|\Tilde{\mathcal{X}}|-1$ continuous functions suffice to preserve $H(\Tilde{X})$, and other four more elements are necessary for preserving the conditional entropies $H(X|U)$, $H(\Tilde{X}|U)$, $H(Y|U)$ and $H(Z|U)$. See \cite[Appendix A]{gunlu2018} for a detailed discussion. Now the converse proof for the CS model is attained.
\qed

\subsection{Achievability Proof}

Similar to the proof of converse part, we begin with the GS model and then argue the CS model.

\medskip
\noindent{\bf \em Achievability Proof of the GS Model}:~~~~ In the proof, we provide only the main contribution of this part, which is the analysis of the privacy-leakage rate for the GS models, since other constraints can be proved by techniques {developed} in previous studies. {As we have seen in the reduction of {\eqref{rlconverse}}, the {maximum} lower bound {on} the privacy-leakage rate for less noisy ACs decreases compared to that of general ACs, so the important objective in the analysis is to check whether the decreased bound can be achieved or not}. For {the} proof of {the} CS model, {it} follows similarly to that of {the} GS model with one-time pad operation to conceal the secret key, and thus it is omitted. The readers may refer to \cite[Appendix A]{vyo2022itw} for a detailed discussion.

Fix the test channel $P_{U|\Tilde{X}}$ and let $\gamma$ be small enough positive. Set $R_S = I(Y;U) - I(Z;U) - 6\gamma$, $R_J = I(\Tilde{X};U|Y) + 4\gamma$, and $R_L = I(X;U|Y) + I(X;Z) + {7\gamma}$, and the sizes of {the} set of helpers $|\mathcal{J}_n| = \exp\{nR_J\}$ and the set of secret keys $|\mathcal{S}_n| = \exp\{nR_S\}$. Define the sets
\begin{align*}
    \mathcal{T}_n &= \Big\{(u^n,\Tilde{x}^n): \frac{1}{n}\log \frac{P_{U^n|\Tilde{X}^n}(u^n|\Tilde{x}^n)}{P_{U^n}(u^n)} \le I(\Tilde{X};U) + \gamma\Big\}, \nonumber \\
    \mathcal{A}_n &= \Big\{(u^n,y^n): \frac{1}{n}\log \frac{P_{Y^n|U^n}(y^n|u^n)}{P_{Y^n}(y^n)} \ge I(Y;U) - \gamma\Big\}, \nonumber \\
    %\mathcal{B}_n &= \Big\{(u^n,x^n,z^n): \nonumber \\
    %&~~~~~~\frac{1}{n}\log \frac{P_{X^n|U^nZ^n}(x^n|u^n,z^n)}{P_{X^n|Z^n}(x^n|z^n)} \ge I(X;U|Z) - \gamma\Big\}, \nonumber \\
    \mathcal{K}_n &= \Big\{(u^n,\Tilde{x}^n,x^n): \nonumber \\
    &~~~~~~\frac{1}{n}\log \frac{P_{\Tilde{X}^n|U^nX^n}(\Tilde{x}^n|u^n,x^n)}{P_{\Tilde{X}^n|X^n}(\Tilde{x}^n|x^n)} \ge I(\Tilde{X};U|X) - \gamma\Big\},
\end{align*}
where $U^n \sim \prod_{t=1}^nP_{U_t}$ with $P_{U_t} = P_U$ for $t \in [1:n]$.

Next, we determine the codebook, and the enrollment (encoding) and authentication (decoding) procedures.
%and we set the size $|\mathcal{Q}_n| = \exp\{n(I(\Tilde{X};U) + 2\gamma)\}$.

\medskip
\noindent{\em Random Code Generation}:~~~Generate $\exp\{n(I(\Tilde{X};U) + 2\gamma)\}$ i.i.d.\ sequences of $\Tilde{u}^n$ from $P_{U}$ and denote the set of these sequences as $\mathcal{Q}_n$. Let $g_n: \Tilde{\mathcal{X}}^n \rightarrow \mathcal{Q}_n\subset \mathcal{U}^n$ be the mapping of measurement $\Tilde{x}^n$ into $\Tilde{u}^n$. The mapping rule is that it searches $\Tilde{u}^n$ such that $(\Tilde{u}^n,\Tilde{x}^n) \in \mathcal{T}_n$. In case there are multiple such $\Tilde{u}^n$, the encoder picks one at random. On the contrary, if there does not exist such a sequence, $\Tilde{u}^n_1$ is chosen. Now prepare $M_J = e^{nR_J}$ bins. Assign each sequence $\Tilde{u}^n \in \mathcal{Q}_n$ to one of $M_J$ bins according to a uniform distribution on $\mathcal{J}_n$. This random assignment is denoted by $\phi_n(\Tilde{u}^n)$. Let $j=\phi_n(\Tilde{u}^n)$, $j \in \mathcal{J}_n$, denote the  bin's index to which $\Tilde{u}^n$ belongs. Also, let $\mathcal{F}_n$ be a universal hash family of functions \cite{carter1979} from $\mathcal{Q}_n$ to $\mathcal{S}_n$. A function $f_n: \mathcal{Q}_n \rightarrow \mathcal{S}_n$ is selected uniformly from $\mathcal{F}_n$ and {satisfies} that $P_{F_n}(\{f_n \in \mathcal{F}_n: f_n(\Tilde{u}^n) = f_n( \widehat{u}^n)\}) \le \frac{1}{|\mathcal{S}_n|}$ for any distinct sequences $\Tilde{u}^n \in \mathcal{Q}_n$ and $ \widehat{u}^n \in \mathcal{Q}_n$, where $P_{F_n}$ is a uniform distribution on $\mathcal{F}_n$.

In the actual encoding and decoding processes, the set $\mathcal{Q}_n$ and the random functions $\phi_n$ and $f_n$ are fixed.

\medskip
\noindent{\em Encoding}:~~~Observing $\Tilde{x}^n$, the encoder first uses $g_n$ to map this sequence to $\Tilde{u}^n \in \mathcal{Q}_n$. It then determines the index $j$ of the bin to which $\Tilde{u}^n$ belongs, i.e., $j = \phi_n(\Tilde{u}^n)$, and generates a secret key $s = f_n(\Tilde{u}^n)$. The index $j$ is shared with the decoder for authentication. 

\medskip
\noindent{\em Decoding}:~~~Seeing $y^n$, the decoder looks for a unique $\widehat{u}^n$ such as $j = \phi_n(\widehat{u}^n)$ and $(\widehat{u}^n,y^n) \in \mathcal{A}_n$. If such a $\widehat{u}^n$ is found, then the decoder {sets} $\psi_n(j,y^n) = \widehat{u}^n$, and distills the secret key $\hat{s} = f_n(\widehat{u}^n)$. Otherwise, the decoder outputs $\hat{s}=f_n(\Tilde{u}^n_1)$ and error is declared.

\medskip
The random codebook $\mathcal{C}_n$ consists of the set  $\mathcal{Q}_n = \{U^n_i : i \in [1: \exp\{n(I(\Tilde{X};U) + 2\gamma)\}]\}$ and the functions $(g_n,\phi_n,\psi_n,f_n)$, and it is revealed to all parties.

By {a} similar argument for evaluating {the} error probability for {the} Wyner-Ziv problem for general sources in \cite{iwatamura2002}, the error probability of the authentication systems averaged over the random codebook vanishes for large enough $n$. The bound on the storage rate is straightforward from the rate setting. The secret-key rate can be proved via \cite[Lemma 3]{Naito2008}, and using \cite[Lemma 12]{wataoha2010} and \cite[Lemma 3]{Naito2008} together, the secrecy-leakage can be made negligible for large enough $n$.

In the rest of this proof, we evaluate the averaged performance of the privacy-leakage rate \eqref{privacy} over all possible $\mathcal{C}_n$. Before diving into the detailed analysis, we introduce some important lemmas for the analysis. This series of lemmas are useful in the proof of the achievability part.

\begin{Lemma} (Iwata and Muramatsu \cite[Lemma 1]{iwatamura2002}) \label{wataoha10}
It holds that
\begin{align}
    \mathbb{E}_{\mathcal{C}_n}[\Pr&\{(g_n(\Tilde{X}^n),Y^n) \notin \mathcal{A}_n~{\rm or}~(g_n(\Tilde{X}^n),X^n,Z^n) \notin \mathcal{B}_n\}] \nonumber \\
    &\le 2\sqrt{\epsilon_n}+\Pr\{(U^n,\Tilde{X}^n) \notin \mathcal{T}_n\} + \exp\{-e^{n\gamma}\}, \label{3939393first}
\end{align}
where
$
\epsilon_n = \Pr\{(U^n,Y^n) \notin \mathcal{A}_n~{\rm or}~(U^n,X^n,Z^n) \notin \mathcal{B}_n\}
$, and where $\mathbb{E}_{\mathcal{C}_n}[\cdot]$ denotes the expectation over the random codebook $\mathcal{C}_n$.
\qed
\end{Lemma}
%Lemma \ref{wataoha10} is an extended version of Wyner's lemma \cite{wyner1975}, and {is} also known as the Markov lemma.
For detailed discussions of the above lemma, the readers should check the proof of \cite[Lemma 1]{iwatamura2002}. By the definitions of $\mathcal{T}_n$, $\mathcal{A}_n$, and $\mathcal{B}_n$, it can be shown that the terms $\Pr\{(U^n,\Tilde{X}^n) \notin \mathcal{T}_n\}$ and $\epsilon_n$ decay exponentially.

\begin{Lemma} (Iwata and Muramatsu \cite{iwatamura2002}) \label{iwatamura}
The error probability averaged over the random assemble is bounded by 
\begin{align}
    \mathbb{E}_{\mathcal{C}_n}&[\Pr\{g_n(\Tilde{X}^n) \neq \psi_n({\phi_n(g_n(\Tilde{X}^n))},Y^n)\}] \nonumber \\
    &\le e^{-\gamma n} + \mathbb{E}_{\mathcal{C}_n}[\Pr\{(g_n(\Tilde{X}^n),Y^n) \notin \mathcal{A}_n\}]. \label{iwatamura11}
\end{align}
%where $\mathbb{E}_{\mathcal{C}_n}[\cdot]$ denotes the expectation over the random codebook $\mathcal{C}_n$.
\qed
\end{Lemma}
%\noindent{{Due to Lemma \ref{wataoha10}, the right-hand side of \eqref{iwatamura11} goes to zero exponentially.}}
Using Lemma \ref{wataoha10}, the right-hand side of \eqref{iwatamura11} converges to zero exponentially. The proof of Lemma \ref{iwatamura} is seen in the argument of evaluating the error probability for the Wyner-Ziv problem with general sources in \cite{iwatamura2002}.
%Now let the pair $(J,S)$ denote the RVs corresponding to the index pair $(j,s)$ of sequence $U^n$ determined by the encoder for $\Tilde{X}^n$. 

Next, we define a security measure
\begin{align*}
    \mu_n = \sum_{z^n \in \mathcal{Z}^n}P_{Z^n}(z^n)\|P_{SJ|Z^n=z^n,\mathcal{C}_n}-P_{\Tilde{S}}P_{J|Z^n=z^n,\mathcal{C}_n}\|,
\end{align*}
where {$\|P_A-P_B\|$ and $P_{\Tilde{S}}$ denote the variational distance between probability distributions $P_A$ and $P_B$}, and the uniform distribution on the set $\mathcal{S}_n$, respectively.

\begin{Lemma} (Watanabe and Oohama \cite[Lemma 12]{wataoha2010}) \label{two}
{An} upper bound of the measure $\mu_n$ averaged over the random codebook is given by
\begin{align}
    \mathbb{E}_{\mathcal{C}_n}[\mu_n] \le e^{-\frac{n\gamma}{2}} + 2\mathbb{E}_{\mathcal{C}_n}[\Pr\{(g_n(\Tilde{X}^n),X^n,Z^n) \notin \mathcal{B}_n\}]. \label{434343}
\end{align}
\qed
\end{Lemma}
%The secret key is generated {and} reproduced by using the function $f_n$, i.e., $S = f_n(g_n(\Tilde{X}^n))$ {and} $\widehat{S} = f_n(\psi_n({J,\hat{Y}^n}))$. %{Similar to the proof in \cite{wataoha2010}, it suffices to pick $f_n$ as a member of a two-universal hash family \cite{carter1979}} to prove the security requirement of \eqref{434343}.
Note that the right-hand side of \eqref{434343} {decays} {exponentially} since so does the second term by using Lemma \ref{wataoha10}.

\begin{Lemma} (Naito et al.\ \cite[Lemma 3]{Naito2008}) \label{naito}
The following conditional entropy is lower bounded {as}
\begin{align}
    H&(S|J,Z^n,\mathcal{C}_n) \nonumber \\
    &\ge (1-\mathbb{E}_{\mathcal{C}_n}[\mu_n])\log M_S + \mathbb{E}_{\mathcal{C}_n}[\mu_n]\log\mathbb{E}_{\mathcal{C}_n}[\mu_n].
\end{align}
\qed
\end{Lemma}

\begin{Lemma} \label{lemma1}
It holds that
\begin{align}
    \mathbb{E}_{\mathcal{C}_n}[\Pr\{(g_n(\Tilde{X}^n),\Tilde{X}^n,X^n) \notin \mathcal{K}_n\}] \le \gamma \label{3939393}
\end{align}
%\begin{align}
%    \mathbb{E}_{\mathcal{C}_n}&[\Pr\{(g_n(\Tilde{X}^n),\Tilde{X}^n,X^n) \notin \mathcal{K}_n\}] \nonumber \\
%    &\le 2\sqrt{\epsilon_n}+\Pr\{(U^n,\Tilde{X}^n) \notin \mathcal{T}_n\} + \exp\{-e^{n\gamma}\}, \label{3939393}
%\end{align}
%where $\epsilon_n = \Pr\{(U^n,\Tilde{X}^n,X^n) \notin \mathcal{K}_n\}$,
for large enough $n$.
%, where $\mathbb{E}_{\mathcal{C}_n}[\cdot]$ denotes the expectation over the random codebook $\mathcal{C}_n$.
\qed
\end{Lemma}
%Lemma \ref{wataoha10} is an extended version of Wyner's lemma \cite{wyner1975}, and {is} also known as the Markov lemma.
%By the definition of the set $\mathcal{K}_n$, the probability $\Pr\{(U^n,\Tilde{X}^n,X^n) \notin \mathcal{K}_n\} \rightarrow 0$ for large enough $n$, and therefore

\begin{Lemma} \label{lemma1-error}
It holds that
\begin{align}
    \mathbb{E}_{\mathcal{C}_n}[\Pr\{(g_n(\Tilde{X}^n),\Tilde{X}^n,X^n) \notin \mathcal{K}_n\}] \le \gamma \label{3939393-1}
\end{align}
%\begin{align}
%    \mathbb{E}_{\mathcal{C}_n}&[\Pr\{(g_n(\Tilde{X}^n),\Tilde{X}^n,X^n) \notin \mathcal{K}_n\}] \nonumber \\
%    &\le 2\sqrt{\epsilon_n}+\Pr\{(U^n,\Tilde{X}^n) \notin \mathcal{T}_n\} + \exp\{-e^{n\gamma}\}, \label{3939393}
%\end{align}
%where $\epsilon_n = \Pr\{(U^n,\Tilde{X}^n,X^n) \notin \mathcal{K}_n\}$,
for large enough $n$, where $\mathbb{E}_{\mathcal{C}_n}[\cdot]$ denotes the expectation over the random codebook $\mathcal{C}_n$.
\qed
\end{Lemma}
%Lemma \ref{wataoha10} is an extended version of Wyner's lemma \cite{wyner1975}, and {is} also known as the Markov lemma.
By the definition of the set $\mathcal{K}_n$, the probability $\Pr\{(U^n,\Tilde{X}^n,X^n) \notin \mathcal{K}_n\} \rightarrow 0$ for large enough $n$, and therefore using \cite[Lemma 1]{iwatamura2002} guarantees that \eqref{3939393} holds. For detailed discussions of the above lemma, the readers may refer to the appendix in \cite{iwatamura2002}.
%By the definitions of $\mathcal{T}_n$ and $\mathcal{K}_n$, it can be shown that the terms $\Pr\{(U^n,\Tilde{X}^n) \notin \mathcal{T}_n\}$ and $\epsilon_n$ decay exponentially.

The following lemma is needed for the analysis of the privacy-leakage rate. The lemma was proved in \cite[Lemma 4]{kitti2015} for a strong typicality set \cite{GK} and \cite[Lemma A4]{vy2} for a modified-weak typicality set \cite{itw3}. Here, a different proof, based on the information-spectrum methods, is given.

\begin{Lemma} \label{kaka} We have that
\begin{align}
H(\Tilde{X}^n|X^n,g_n(\Tilde{X}^n),\mathcal{C}_n) \le n(H(\Tilde{X}|X,U) + {2\gamma} + r_n), \label{kaka1}
\end{align}
where
$r_n = {1/n(1 - \log(1-\gamma))} + {\gamma}\log |\Tilde{\mathcal{X}}|$, and $r_n$ {tends} to zero as $n$ {approaches} infinity {and $\gamma \downarrow 0$}.
\end{Lemma}
\noindent{P}roof:~~~~The proof is given in Appendix B.
\qed

\medskip
Using Lemma \ref{iwatamura}, the ensemble average of the error probability of encoding and decoding can be made exponentially vanish for large enough $n$. Also, the bound on the storage rate is straightforward from the rate setting.

\medskip
\noindent{\em Analysis of Secret-Key Rate}:~~~~Using Lemma \ref{naito}, it holds that
\begin{align}
    H(S|\mathcal{C}_n) &\ge H(S|J,Z^n,\mathcal{C}_n) \nonumber \\
    &\ge (1-\mathbb{E}_{\mathcal{C}_n}[\mu_n])\log M_S + \mathbb{E}_{\mathcal{C}_n}[\mu_n]\log \mathbb{E}_{\mathcal{C}_n}[\mu_n] \nonumber \\
    &= (1-\mathbb{E}_{\mathcal{C}_n}[\mu_n]) nR_S + \mathbb{E}_{\mathcal{C}_n}[\mu_n]\log \mathbb{E}_{\mathcal{C}_n}[\mu_n] \nonumber \\
    &= n(R_S - \mathbb{E}_{\mathcal{C}_n}[\mu_n](R_S - \frac{1}{n}\log \mathbb{E}_{\mathcal{C}_n}[\mu_n])) \nonumber \\
    &\ge n(R_S - \gamma) \label{hsrs111}
\end{align}
for large enough $n$ because $\mathbb{E}_{\mathcal{C}_n}[\mu_n]$ decays exponentially (cf. Lemma \ref{two}).

\medskip
\noindent{\em Analysis of Secrecy-Leakage}:~~~~We can {expand} the left-hand side of \eqref{secrecy} as
\begin{align}
    I(S;J,Z^n|\mathcal{C}_n) &= H(S|\mathcal{C}_n) - H(S|J,Z^n,\mathcal{C}_n) \nonumber \\
    &\le \log M_S - H(S|J,Z^n,\mathcal{C}_n) \nonumber \\
    %&\le \log M_S - H(S|J,Z^n,F,\mathcal{C}_n) \nonumber \\
    &\overset{\rm (a)}\le \mathbb{E}_{\mathcal{C}_n}[\mu_n](\log M_S - \log\mathbb{E}_{\mathcal{C}_n}[\mu_n]) \nonumber \\
    &\le \gamma \label{isjzn123}
\end{align}
{for sufficiently large $n$}, where (a) follows from Lemma \ref{naito} and the last inequality is due to Lemma \ref{two}.

\medskip
\noindent{\em Analysis of Privacy-Leakage {Rate}}:~~~~For \eqref{privacy}, we have that
\vspace{-1mm}
\begin{align}
    I(X^n;J,Z^n&|\mathcal{C}_n) = I(X^n;J|\mathcal{C}_n) + I(X^n;Z^n|J,\mathcal{C}_n) \nonumber \\
    &\overset{\mathrm{(a)}}= I(X^n;J|\mathcal{C}_n) + H(Z^n|J,\mathcal{C}_n) - H(Z^n|X^n) \nonumber \\
    &\overset{\mathrm{(b)}}\le I(X^n;J|\mathcal{C}_n) + nI(X;Z), \label{privacylast}
\end{align}
where (a) holds because for a given $\mathcal{C}_n$, the Markov chain $J-X^n-Z^n$ holds, and $(X^n,Z^n)$ are independent of $\mathcal{C}_n$, and (b) follows because conditioning reduces entropy.

Next, we focus on bounding the term $I(X^n;J|\mathcal{C}_n)$ in \eqref{privacylast}:
\begin{align*}
    I&(X^n;J|\mathcal{C}_n) = H(J|\mathcal{C}_n) - H(J|X^n,\mathcal{C}_n) \nonumber \\
    &\le nR_J - H(J|X^n,\mathcal{C}_n) \nonumber \\
    &= nR_J - H(\Tilde{X}^n,J|X^n,\mathcal{C}_n) + H(\Tilde{X}^n|X^n,J,\mathcal{C}_n) \nonumber \\
    &\overset{\mathrm{(c)}}\le nR_J - nH(\Tilde{X}|X) + H(\Tilde{X}^n|X^n,J,\mathcal{C}_n) \nonumber \\
    &\overset{\mathrm{(d)}}= nR_J - nH(\Tilde{X}|X) + H(\Tilde{X}^n|X^n,g_n(\Tilde{X}^n),\mathcal{C}_n) \nonumber \\
    &~~~+I(\Tilde{X}^n;g_n(\Tilde{X}^n)|X^n,J,\mathcal{C}_n) \nonumber \\
    &=  nR_J - nH(\Tilde{X}|X) + H(\Tilde{X}^n|X^n,g_n(\Tilde{X}^n),\mathcal{C}_n) \nonumber \\
    &~~~+H(g_n(\Tilde{X}^n)|X^n,J,\mathcal{C}_n)
\end{align*}
\begin{align}
    &\overset{\mathrm{(e)}}=  nR_J - nH(\Tilde{X}|X) + H(\Tilde{X}^n|X^n,g_n(\Tilde{X}^n),\mathcal{C}_n) \nonumber \\
    &~~~+H(g_n(\Tilde{X}^n)|X^n,J,Y^n,\mathcal{C}_n) \nonumber \\
    &\overset{\mathrm{(f)}}\le  nR_J - nH(\Tilde{X}|X) + H({\Tilde{X}^n}|X^n,g_n(\Tilde{X}^n),\mathcal{C}_n) \nonumber \\
    &~~~+H(g_n(\Tilde{X}^n)|J,Y^n,\mathcal{C}_n) \nonumber \\
    &\overset{\mathrm{(g)}}\le  nR_J - nH(\Tilde{X}|X) + H(\Tilde{X}^n|X^n,g_n(\Tilde{X}^n),\mathcal{C}_n) + n\delta_n \nonumber \\
    &\overset{\mathrm{(h)}}\le  n(R_J - H(\Tilde{X}|X) + H({\Tilde{X}}|X,U) + {2\gamma} + r_n + \delta_n) \nonumber \\
    &=  n(R_J - I(\Tilde{X};U|X) + {2\gamma} + r_n + \delta_n) \nonumber \\
    &\overset{\mathrm{(i)}}=  n(R_J - (I(\Tilde{X};U)-I(X;U)) + {2\gamma} + r_n + \delta_n) \nonumber \\
    &=  n(I(X;U)-I(Y;U) + {6\gamma} + r_n + \delta_n), \label{asas}
\end{align}
where (c) holds as $(\Tilde{X}^n,X^n)$ are independent of $\mathcal{C}_n$, (d) follows because $J$ is a function of $g_n(\Tilde{X}^n)$, i.e., $J = \phi(g_n(\Tilde{X}^n))$, (e) is due to the Markov chain $g_n(\Tilde{X}^n)-(X^n,J)-Y^n$, (f) follows because conditioning reduces entropy, (g) follows as the codeword $g_n(\Tilde{X}^n)$ can be estimated from $(J,Y^n)$ with high probability, and thus Fano's inequality is applied, (h) follows from Lemma \ref{kaka}, (i) is due to the Markov chain $U-\Tilde{X}-X$, and the last equality holds as we set $R_J = I(\Tilde{X};U|Y) + 4\gamma$.

Merging \eqref{privacylast} and \eqref{asas}, we obtain that
\begin{align}
    I(X^n;J,Z^n|\mathcal{C}_n) &\le n(I(X;U|Y) + I(X;Z) + {8\gamma}) \nonumber \\
    &= n(R_L + \gamma) \label{rllast1111}
\end{align}
for large enough $n$, {which gives the desired bound on the privacy-leakage rate constraint \eqref{privacy} in Definition \ref{def1}, and this also hints} that the decreased lower bound {on} the privacy-leakage rate in \eqref{rlconverse} is achievable.

Finally, applying the selection lemma \cite[Lemma 2.2]{BB}, there exists at least one good codebook that satisfies all conditions in Definition \ref{def1}.

\begin{figure}[!t]
    \centering
    \includegraphics[scale=0.6]{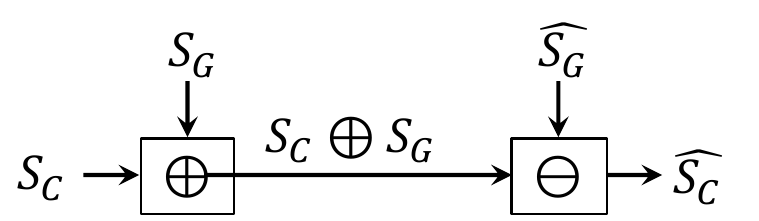}
    \caption{One-time operation to mask the secret key in CS model for secure transmission}
    \label{sgsc}
    \vspace{-5mm}
\end{figure}

\medskip
\noindent{\bf \em Achievability Proof of the CS Model}:~~~~
The proof of this model follows similarly to that of the GS model with an additional one-time pad operation to conceal the secret-key for secure transmission. See Fig.\ \ref{sgsc} for an explanation of detailed procedures. In the following discussion, let $(J_G,S_G)$ and $(J_C,S_C)$ denote the pairs of helper data and secret key for the GS and CS models, respectively. Here, the operators $\oplus$ and $\ominus$ stand for addition and subtraction modulo $|\mathcal{S}_n|$. The encoder and decoder of the GS model are used as components inside the encoder and decoder of the CS model.
%These components obey the encoding and decoding schemes of GS model. The rate settings of the components also the same as in the proof of the GS model.
Also, the components and the encoder and decoder share the same codebook.

\medskip
\noindent{\em Encoding}:~~~
Observing $\Tilde{X}^n$ and $S_C$, the encoder uses the component to determines $S_G$ and then masks $S_C$ with the generated key $S_G$. After that, it combines the masked information $S_C \oplus S_G$ with the helper data $J_G$ to form $J_C = (J_G,S_C \oplus S_G)$, and stores this information in a public DB for future authentication.

\medskip
\noindent{\em Decoding}:~~~~
Seeing $Y^n$, the decoder uses the component to estimate $\hat{S}_G$, and this estimation is utilized to reconstruct the secret key $\hat{S}_C$ by subtracting the second half of the helper data $J_C$, i.e., $S_C \oplus S_G$, with $\hat{S}_G$. That is,
\begin{align}
    \hat{S}_C = S_C \oplus S_G \ominus \hat{S}_G. \label{sgsc11}
\end{align}

\medskip
\noindent{\em Performance Analyses}:~~~~For error probability of the CS model, \eqref{sgsc11} indicates that $\hat{S}_C = S_C$ if and only if $\hat{S}_G = S_G$. This means $\Pr\{\hat{S}_C \neq S_C\} =\Pr\{ \hat{S}_G \neq S_G\}$. Using Lemma \ref{iwatamura}, it guarantees that the error probability of the CS model can be made negligible for large enough $n$. The bound on the secret-key rate follows directly from the rate settings since it is chosen uniformly from the set $\mathcal{S}_n$. For the storage rate, we have that
\begin{align}
    &{\rm Total~storage~rate} \le \frac{1}{n}\log M_J + \frac{1}{n}\log M_S \nonumber \\
    &~~~~~~~~~~~~~~= I(\Tilde{X};U|Y) + 4 \delta + I(Y;U) - I(Z;U) - 6\delta \nonumber \\
    &~~~~~~~~~~~~~~\le I(\Tilde{X};U|Z), \label{storagecs111}
\end{align}
where the last equality holds due to the Markov chains $U-\Tilde{X}-Y$ and $U-\Tilde{X}-Z$.

For the secrecy-leakage, by a similar argument of the development in \cite[Secrecy-leakage Rate]{gksc2018}, we have that $I(S_C;J_C,Z^n|\mathcal{C}_n) \le \log M_S - H(S_G|\mathcal{C}_n) + I(S_G;J_G,Z^n|\mathcal{C}_n)$. Using \eqref{hsrs111} and \eqref{isjzn123}, it follows that
\begin{align}
    \frac{1}{n},I(S_C;J_C,Z^n|\mathcal{C}_n) \le 2\gamma \label{secrecycsss}
\end{align}
for large enough $n$.

The privacy-leakage can be evaluated as follows. It holds that
\begin{align} 
    I(X^n;J_C,Z^n|\mathcal{C}_n) = I(X^n;J_G,Z^n|\mathcal{C}_n). \label{ixn111}
\end{align}
First, observe that
\begin{align}
    &I(X^n;J_C,Z^n|\mathcal{C}_n) = I(X^n;J_G,S_C\oplus S_G,Z^n|\mathcal{C}_n) \nonumber \\
    &= I(X^n;J_G,Z^n|\mathcal{C}_n) + I(X^n;S_C\oplus S_G|J_G,Z^n,\mathcal{C}_n). \label{ixn112}
\end{align}
Next, we bound the second term in \eqref{ixn112}.
\begin{align}
    &I(X^n;S_C\oplus S_G|J_G,Z^n,\mathcal{C}_n) \nonumber \\
    &= H(S_C\oplus S_G|J_G,Z^n,\mathcal{C}_n) - H(S_C\oplus S_G|J_G,Z^n,X^n,\mathcal{C}_n) \nonumber \\
    &\le \log M_S - H(S_C\oplus S_G|J_G,Z^n,X^n,\mathcal{C}_n) \nonumber \\
    &\overset{\mathrm{(a)}}\le \log M_S - H(S_C\oplus S_G|J_G,Z^n,X^n,S_G,\mathcal{C}_n) \nonumber \\
    &\le \log M_S - H(S_C|J_G,Z^n,X^n,S_G,\mathcal{C}_n) \nonumber \\
    &\overset{\mathrm{(b)}}= \log M_S - H(S_C) \nonumber \\
    &\overset{\mathrm{(c)}}= 0, \label{ixn113}
\end{align}
where (a) follows because conditioning does not increase entropy, (b) is because the tuple $(J_G,Z^n,X^n,S_G)$ is independent of $S_C$, and (c) holds because $S_C$ is independent of the codebook $\mathcal{C}_n$ and chosen uniformly from the set $\mathcal{S}_n$.

Since mutual information takes non-negative values, from \eqref{ixn113}, it implies that $I(X^n;S_C\oplus S_G|J_G,Z^n,\mathcal{C}_n) = 0$. Due to this fact, \eqref{ixn111} holds.
Now invoking \eqref{rllast1111} and from \eqref{ixn111}, we have that
\begin{align}
    I(X^n;J_C,Z^n|\mathcal{C}_n) \le n(R_L + \gamma) \label{privacy-leakagecsss}
\end{align}
for large enough $n$.

Applying the selection lemma \cite[Lemma 2.2]{BB} to Equations, e.g., \eqref{storagecs111}, \eqref{secrecycsss}, \eqref{privacy-leakagecsss}, there exists at least one good codebook satisfying all the conditions in Definition \ref{def2}.
\qed

\vspace{-2mm}

\section{Proof of Lemma \ref{kaka}}
Define a binary {random variable} $T = \mathrm{\bf 1}\{(g_n(\Tilde{X}^n),\Tilde{X}^n,X^n) \in \mathcal{K}_n\}$, where $\mathrm{\bf 1}\{\cdot\}$ denotes {the} indicator function. When $T=0$, using Lemma \ref{lemma1}, it is straightforward that
$
    \mathbb{E}_{\mathcal{C}_n}[P_{T}(0)] \le \gamma.
$
%and the probability when $T=1$ is given by
%\begin{align}
%    \mathbb{E}_{\mathcal{C}_n}[P_T(1)]=\mathbb{E}_{\mathcal{C}_n}\left[\sum_{(\Tilde{x}^n,x^n) \in \mathcal{K}_n\cap\mathcal{L}_n}P_{\Tilde{X}^nX^ng_n(\Tilde{X}^n)}(\Tilde{X}^nX^ng_n(\Tilde{X}^n))\right]. \label{tis1}
%\end{align}

In the rest of equation developments, let $c_n$ and $\Tilde{u}^n$ be realizations of the random codebook $\mathcal{C}_n$ and the mapping function $g_n(\Tilde{X}^n)$, namely, $\Tilde{u}^n = g_n(\Tilde{x}^n)$, respectively. The conditional entropy on the left-hand side of \eqref{kaka1} can be evaluated as
\vspace{-2mm}
\begin{align}
    &H(\Tilde{X}^n|X^n,g_n(\Tilde{X}^n),\mathcal{C}_n)
    \le H(\Tilde{X}^n,T|X^n,g_n(\Tilde{X}^n),\mathcal{C}_n) \nonumber \\
    &\le H(T) + H(\Tilde{X}^n|X^n,g_n(\Tilde{X}^n),T,\mathcal{C}_n) \nonumber \\
    %&\le 1 + \sum_{c_n \in \mathcal{C}_n}P_{\mathcal{C}_n}(c_n)H(\Tilde{X}^n|X^n,g_n(\Tilde{X}^n),T,\mathcal{C}_n=c_n) \nonumber \\
    &\le 1 + \mathbb{E}_{\mathcal{C}_n}[P_{T}(0)]H(\Tilde{X}^n) \nonumber \\
    &~+ \sum_{c_n}P_{T,\mathcal{C}_n}(1,c_n)H(\Tilde{X}^n|X^n,g_n(\Tilde{X}^n),T=1,\mathcal{C}_n=c_n) \nonumber \\
    &\le 1 + n\gamma\log|\Tilde{\mathcal{X}}| \nonumber \\
    &~+ \sum_{c_n}P_{T,\mathcal{C}_n}(1,c_n)H(\Tilde{X}^n|X^n,g_n(\Tilde{X}^n),T=1,\mathcal{C}_n=c_n), \label{corona}
\end{align}
where the last inequality is due to Lemma \ref{lemma1}. Next, we concentrate only on bounding the conditional entropy $H(\Tilde{X}^n|X^n,g_n(\Tilde{X}^n),T=1,\mathcal{C}_n=c_n)$ in \eqref{corona}. For a given $\mathcal{C}_n = c_n$, we define the following probability distribution
\begin{align}
&P_{g_n(\Tilde{X}^n)\Tilde{X}^nX^n|T}(\Tilde{u}^n,\Tilde{x}^n,x^n|1) \nonumber \\
&= \begin{cases}
\frac{P_{g_n(\Tilde{X}^n)\Tilde{X}^nX^n}(\Tilde{u}^n,\Tilde{x}^n,x^n)}{P_T(1)} &~\mathrm{if}~(\Tilde{u}^n,\Tilde{x}^n,x^n) \in \mathcal{K}_n \\
0 &~\mathrm{otherwise}
\end{cases}, \label{cases}
\end{align}
and
$
    P_T(1) = \sum_{(\Tilde{u}^n,\Tilde{x}^n,x^n) \in \mathcal{K}_n}P_{g_n(\Tilde{X}^n)\Tilde{X}^nX^n}(\Tilde{u}^n,\Tilde{x}^n,x^n),
$
which is obvious from the definition of the random variable $T$.
For every tuple $(\Tilde{u}^n,\Tilde{x}^n,x^n) \in \mathcal{K}_n$, observe that
\begin{align}
    &P_{\Tilde{X}^n|X^ng_n(\Tilde{X}^n)T}(\Tilde{x}^n|x^n,\Tilde{u}^n,1)
    %&~~~~= \frac{P_{g_n(\Tilde{X}^n)\Tilde{X}^nX^n|T}(g_n(\Tilde{x}^n),\Tilde{x}^n,x^n|1)}{P_{X^ng_n(\Tilde{X}^n)|T}(x^n,g_n(\Tilde{x}^n)|1)} \nonumber \\
    %&~~~~= \frac{P_T(1)P_{g_n(\Tilde{X}^n)\Tilde{X}^nX^n)|T}(g_n(\Tilde{x}^n),\Tilde{x}^n,x^n|1)}{P_{X^ng_n(\Tilde{X}^n)T}(x^n,g_n(\Tilde{x}^n),1)} \nonumber \\
    \overset{\mathrm{(a)}}= \frac{P_{g_n(\Tilde{X}^n)\Tilde{X}^nX^n}(\Tilde{u}^n,\Tilde{x}^n,x^n)}{P_{g_n(\Tilde{X}^n)X^nT}(\Tilde{u}^n,x^n,1)} \nonumber \\
    &~\ge \frac{P_{g_n(\Tilde{X}^n)\Tilde{X}^nX^n}(\Tilde{u}^n,\Tilde{x}^n,x^n)}{P_{g_n(\Tilde{X}^n)X^n}(\Tilde{u}^n,x^n)} = P_{\Tilde{X}^n|g_n(\Tilde{X}^n)X^n}(\Tilde{x}^n|\Tilde{u}^n,x^n), \label{ee333}
\end{align}
where (a) is due to \eqref{cases}. Also, we have that
\begin{align}
    &\log\frac{1}{P_{\Tilde{X}^n|X^ng_n(\Tilde{X}^n)}(\Tilde{x}^n|x^n,\Tilde{u}^n)} \nonumber \\
    &= \log\frac{P_{\Tilde{X}^n|X^nU^n}(\Tilde{x}^n|x^n,\Tilde{u}^n)}{P_{\Tilde{X}^n|X^ng_n(\Tilde{X}^n)}(\Tilde{x}^n|x^n,\Tilde{u}^n)} \nonumber \\
    &~~~~+ \log\frac{P_{\Tilde{X}^n|X^n}(\Tilde{x}^n|x^n)}{P_{\Tilde{X}^n|X^nU^n}(\Tilde{x}^n|x^n,\Tilde{u}^n)} + \log\frac{1}{P_{\Tilde{X}^n|X^n}(\Tilde{x}^n|x^n)} \nonumber \\
    &\overset{\rm (b)}\le \log\frac{P_{\Tilde{X}^n|X^nU^n}(\Tilde{x}^n|x^n,\Tilde{u}^n)}{P_{\Tilde{X}^n|X^ng_n(\Tilde{X}^n)}(\Tilde{x}^n|x^n,\Tilde{u}^n)} \nonumber \\
    &~~~~-n(I(\Tilde{X};U|X) - \gamma) + n(H(\Tilde{X}|X) + \gamma) \nonumber \\
    &= \log\frac{P_{\Tilde{X}^n|X^nU^n}(\Tilde{x}^n|x^n,\Tilde{u}^n)}{P_{\Tilde{X}^n|X^ng_n(\Tilde{X}^n)}(\Tilde{x}^n|x^n,\Tilde{u}^n)} + n(H(\Tilde{X}|X,U) + 2\gamma) \label{hxu2g}
\end{align}
for all large $n$, where (b) follows because the condition of the set $\mathcal{K}_n$ is applied to the second term, and using the law of large numbers, the i.i.d. property of $(\Tilde{X}^n,X^n)$ guarantees that
$
    \log\frac{1}{P_{\Tilde{X}^n|X^n}(\Tilde{x}^n|x^n)} \le n(H(\Tilde{X}|X) + \gamma)
$
for large enough $n$.
%$(g_n(\Tilde{x}^n),\Tilde{x}^n,x^n) \in \mathcal{K}_n\cap\mathcal{L}_n$ and thus, the conditions of sets $\mathcal{K}_n$ and $\mathcal{L}_n$ are applied to the first and second terms, respectively.

In light of \eqref{corona}, we have that
\vspace{-1mm}
\begin{align}
&H(\Tilde{X}^n|X^n,g_n(\Tilde{X}^n),T=1,\mathcal{C}_n=c_n) \nonumber \\
    &\le H(\Tilde{X}^n|X^n,g_n(\Tilde{X}^n),T=1) \nonumber \\
    &= \sum_{\substack{(\Tilde{u}^n,\Tilde{x}^n,x^n) \in \mathcal{K}_n}}P_{g_n(\Tilde{X}^n)\Tilde{X}^nX^nT}(\Tilde{u}^n,\Tilde{x}^n,x^n,1) \nonumber \\
    &~~\cdot\Big(\log \frac{1}{P_{\Tilde{X}^n|X^ng_n(\Tilde{X}^n)T}(\Tilde{x}^n|x^n,\Tilde{u}^n,1)}\Big) \nonumber \\
    &\overset{\mathrm{(c)}}\le \sum_{\substack{(\Tilde{u}^n,\Tilde{x}^n,x^n) \in \mathcal{K}_n}}P_{g_n(\Tilde{X}^n)\Tilde{X}^nX^nT}(\Tilde{u}^n,\Tilde{x}^n,x^n,1)\nonumber \\
    &~~\cdot\Big(\log {\frac{1}{P_{\Tilde{X}^n|X^ng_n(\Tilde{X}^n)}(\Tilde{x}^n|x^n,\Tilde{u}^n)}}\Big) \nonumber \\
    &\overset{\mathrm{(d)}}\le \sum_{\substack{(\Tilde{u}^n,\Tilde{x}^n,x^n) \in \mathcal{K}_n}}P_T(1)\cdot P_{g_n(\Tilde{X}^n)\Tilde{X}^nX^n|T}(\Tilde{u}^n,\Tilde{x}^n,x^n|1)\nonumber \\
    &~~\cdot\Big(\log\frac{P_{\Tilde{X}^n|X^nU^n}(\Tilde{x}^n|x^n,\Tilde{u}^n)}{P_{\Tilde{X}^n|X^ng_n(\Tilde{X}^n)}(\Tilde{x}^n|x^n,\Tilde{u}^n)} + n(H(\Tilde{X}|X,U) + 2\gamma)\Big) \nonumber \\
    %&\le \sum_{\substack{(g(\Tilde{x}^n),\Tilde{x}^n,x^n) \in \mathcal{K}_n\cap\mathcal{L}_n}}P_{g_n(\Tilde{X}^n)\Tilde{X}^nX^n}(g_n(\Tilde{x}^n),\Tilde{x}^n,x^n)\nonumber \\
    %&~~~~\cdot n(H(\Tilde{X}|X,U)+ 2\gamma) \nonumber \\
    &\overset{\mathrm{(e)}}\le n(H(\Tilde{X}|X,U)+ 2\gamma) {- \log(1-\gamma)}, \label{goal}
\end{align}
where (c) and (d) follow from \eqref{ee333} and \eqref{hxu2g}, respectively, and (e) is due to \eqref{first-term1} {shown below}. {To derive} \eqref{first-term1}, we define {$A^n = (g_n(\Tilde{X}^n),\Tilde{X}^n,X^n)$ and $a^n = (\Tilde{u}^n,\Tilde{x}^n,x^n)$ for brevity. {From} \eqref{cases}, it follows that $P_{g(\Tilde{X}^n)\Tilde{X}^nX^n}(\Tilde{u}^n,\Tilde{x}^n,x^n) = P_{T}(1)\cdot P_{A^n|T}(a^n|1)$, {and thus we have that}
\begin{align}
&\sum_{\substack{a^n \in \mathcal{K}_n}}P_{A^n|T}(a^n|1)\log\frac{P_{\Tilde{X}^n|X^nU^n}(\Tilde{x}^n|x^n,\Tilde{u}^n)}{P_{\Tilde{X}^n|X^ng_n(\Tilde{X}^n)}(\Tilde{x}^n|x^n,\Tilde{u}^n)} \nonumber \\
%&\le \sum_{\substack{a^n \in \mathcal{K}_n}}P_{A^n}(a^n)\log\frac{P_{\Tilde{X}^n|X^nU^n}(\Tilde{x}^n|x^n,\Tilde{u}^n)}{P_{\Tilde{X}^n|X^ng_n(\Tilde{X}^n)}(\Tilde{x}^n|x^n,\Tilde{u}^n)} \nonumber \\
&\overset{\mathrm{(f)}}\le \log\Biggl(\sum_{\substack{a^n \in \mathcal{K}_n}}\frac{P_{A^n|T}(a^n|1)\cdot P_{\Tilde{X}^n|X^nU^n}(\Tilde{x}^n|x^n,\Tilde{u}^n)}{P_{\Tilde{X}^n|X^ng_n(\Tilde{X}^n)}(\Tilde{x}^n|x^n,\Tilde{u}^n)}\Biggl) \nonumber \\
&= \log\Biggl(\sum_{\substack{a^n \in \mathcal{K}_n}}\frac{P_{g_n(\Tilde{X}^n)X^n}(\Tilde{u}^n,x^n)\cdot P_{\Tilde{X}^n|X^nU^n}(\Tilde{x}^n|x^n,\Tilde{u}^n)}{P_{T}(1)}\Biggl) \nonumber \\
&\le \log\Biggl(\sum_{\substack{(\Tilde{u}^n,x^n) \in \mathcal{Q}_n\times\mathcal{X}^n}}P_{g_n(\Tilde{X}^n)X^n}(\Tilde{u}^n,x^n) \nonumber \\
&~~~~~~~~~~\cdot \biggl(\sum_{\substack{\Tilde{x}^n \in \mathcal{\Tilde{X}}^n}}P_{\Tilde{X}^n|X^nU^n}(\Tilde{x}^n|x^n,\Tilde{u}^n)\biggl)\Biggl) - \log P_T(1)\nonumber \\
&\overset{\mathrm{(g)}}\le -\log(1-\gamma), \label{first-term1}
\end{align}
where (f) is due to Jensen's inequality and (g) follows because Lemma \ref{lemma1}, implying that $P_T(1) \ge 1 - \gamma$, is used}.

{Lastly,} substituting \eqref{goal} into \eqref{corona}, it follows that
\begin{align}
    H(\Tilde{X}^n|X^n,g_n(\Tilde{X}^n),\mathcal{C}_n) \le n(H(\Tilde{X}|X,U)+ 2\gamma + r_n),
\end{align}
where $r_n = 1/n(1 - \log(1-\gamma)) + \gamma\log |\Tilde{\mathcal{X}}|$.
\qed

\section{Proof of Theorem \ref{th3}}
Note that due to the {uniformity} of the sources, the reverse channel $P_{X|\Tilde{X}}$ also results in a {binary symmetric channel} with crossover probability $p$, {and the entropies $H(\Tilde{X})$, $H(X)$, and $H(Z)$ are all equal to one}.

\medskip
\noindent{\em Achievability}:~~~~We begin by proving the inner region {of} $\mathcal{R}_G$. Observe that each rate constraint in the region can be bounded as follows:
\begin{align}
    R_S &\le I(Y;U) - I(Z;U)\overset{\mathrm{(a)}}= (1-q)I(X;U) - I(Z;U) \nonumber \\
        &= H(Z|U) - (1-q)H(X|U) - q \label{rsbi} \\
        &\overset{\mathrm{(b)}}= H_b(\beta*p*\epsilon) - (1-q)H_b(\beta*p) - q, \label{rsbi1} \\
    R_J & \ge I(\Tilde{X};U|Y) = I(\Tilde{X};U) - I(Y;U) \nonumber \\
        &= I(\Tilde{X};U) - (1-q)I(X;U) \nonumber \\
        &= q + (1-q)H(X|U) - H(\Tilde{X}|U) \label{rjbi} \\
        &\overset{\mathrm{(c)}}= q + (1-q)H_b(\beta*p) - H_b(\beta),\label{rjbi1} \\
    R_L & \ge I(X;U|Y) + {I(X;Z)} = qI(X;U) + I(X;Z) \nonumber \\
        &= 1+q -qH(X|U) - H(X|Z) \label{rlbi} \\
        &\overset{\mathrm{(d)}}= 1+q -qH_b(\beta*p) - {H_b(\epsilon)}, \label{rlbi1}
\end{align}
where (a) follows because $Y = X$ with probability $1-q$, {and (b), (c), and (d) are achieved by considering the test channel $P_{U|\Tilde{X}}$ to be a {binary symmetric channel} with crossover probability $\beta$}. For the CS model, we argue only for the storage rate as the others follow the same analysis seen in {the} GS model:
\begin{align}
    R_J & \ge I(\Tilde{X};U|Z) \overset{\mathrm{(a)}}= I(\Tilde{X};U) - I(Z;U) \nonumber \\
        &= H(Z|U) - H(\Tilde{X}|U) \label{rjcs}\\
        &= H_b(\beta*p*\epsilon) - H_b(\beta) \label{rjcs1},
\end{align}
where (a) follows from the Markov chain $U-\Tilde{X}-Z$.
\qed

\medskip
\noindent{\em Converse Part}:~~~
Before the proof, we introduce a simple lemma that will be used to match the inner and outer bounds of the capacity regions for binary sources.
\begin{Lemma} \label{binarylemma} Given $p \in [0,\frac{1}{2}]$ and $\epsilon \in [0,\frac{1}{2}]$, and for any $\lambda \in [0,1/2]$, it holds that
\begin{align}
    \frac{\lambda*p - \epsilon}{1-2\epsilon} \le \lambda*p*\epsilon \le \frac{1}{2}, \label{lemma55}
\end{align}
\rev{where the special case of $\epsilon = \frac{1}{2}$ for the fraction of the left-hand side in \eqref{lemma55} should be interpreted as $\lim_{\epsilon \rightarrow \frac{1}{2}^-} \frac{\lambda*p - \epsilon}{1-2\epsilon}$}.
\end{Lemma}
\noindent{P}roof:~~~~First, the second inequality in \eqref{lemma55} follows {from} the reason that for given $p$ and $\epsilon$, the function $\lambda*p*\epsilon$ is non-decreasing {with respect to} $\lambda$, and its peak is $1/2$ at the point $\lambda = 1/2$.

Next, the relation of the first inequality in \eqref{lemma55} is verified. We begin by mentioning an extreme case where $\epsilon = \frac{1}{2}$. When $\epsilon$ approaches $\frac{1}{2}$, we consider other two subcases where ($p = \frac{1}{2}$, $\lambda \in [0, \frac{1}{2}]$) and ($p < \frac{1}{2}$, $\lambda \in [0, \frac{1}{2}]$). For the former subcase, the limit value $\lim_{\epsilon \rightarrow {\frac{1}{2}}^-} \frac{\lambda*p - \epsilon}{1-2\epsilon}$ is $\frac{1}{2}$ regardless of $\lambda$. The first inequality in \eqref{lemma55} holds for this subcase as $\lambda*p*\epsilon = \frac{1}{2}$ for $p = \frac{1}{2}$. For {the} latter subcase, we have that
\begin{align}
    \lim_{\epsilon \rightarrow {\frac{1}{2}}^-} \frac{\lambda*p - \epsilon}{1-2\epsilon} = 
    \begin{cases}
-\infty & (\lambda < \frac{1}{2})\\
\frac{1}{2} & (\lambda = \frac{1}{2})
\end{cases}.
\end{align}
Since $\lambda*p*\epsilon \ge 0$ and $\lambda*p*\epsilon = \frac{1}{2}$ at $\lambda = \frac{1}{2}$, the first inequality in \eqref{lemma55} also holds for the latter subcase.

In the remaining of the proof, we focus on the range of $\epsilon \in  [0,\frac{1}{2})$. Due to the same reason {for} the second inequality in \eqref{lemma55}, it is obvious that $\lambda*p \le 1/2$ or ${2(\lambda*p)} \le 1$, and since $\epsilon \ge 0$, it follows that $-2\epsilon(1-\epsilon) \le -4\epsilon(1-\epsilon)(\lambda*p)$. Adding $\lambda*p$ to both sides of this inequality, we have that $-2\epsilon+2\epsilon^2 + \lambda*p \le -4\epsilon(\lambda*p) + 4\epsilon^2(\lambda*p) + \lambda*p$. Rearrange both sides as follows:
\begin{align}
    -\epsilon + \lambda*p &\le \epsilon-2\epsilon^2 - 4\epsilon(\lambda*p) + 4\epsilon^2(\lambda*p) + \lambda*p \nonumber \\
    &= (1-2\epsilon)(\epsilon - 2\epsilon ({\lambda}*p) + \lambda*p) \nonumber \\
    &= (1-2\epsilon)(\lambda*p*\epsilon), \label{4444}
\end{align}
indicating that Lemma \ref{binarylemma} holds.
\qed

\medskip
To derive the outer region of {the} GS model, we need to further bound \eqref{rsbi}, \eqref{rjbi}, and \eqref{rlbi}. In order to do so, we fix $H(X|U)$ and derive tight upper bounds for both $H(Z|U)$ and $H(\Tilde{X}|U)$. Since $H_b(p) = H(X|\Tilde{X}) \le H(X|U) \le H(X) = 1$, we {fix $\lambda \in [0,1/2]$ such that}
\begin{align}
    H(X|U) = H_b(\lambda*p). \label{xubi}
\end{align}
In the direction from $Z$ to $X$, using {Mrs. Gerber’s Lemma} \cite{wyner}, it follows that
\begin{align}
    H(X|U) \ge H_b(H_b^{-1}(H(Z|U))*\epsilon). \label{hxu111}
\end{align}
\rev{Now substituting the value of $H(X|U)$ that we have set in \eqref{xubi} into the left-hand side of \eqref{hxu111}, we have that
\begin{align}
    H_b(\lambda*p) \ge H_b(H_b^{-1}(H(Z|U))*\epsilon).
\end{align}
Since all $\lambda,p,\epsilon \le \frac{1}{2}$ and the binary entropy function is monotonously increasing in the interval $[0,\frac{1}{2}]$, it follows that
\begin{align*}
    \lambda*p \ge H_b^{-1}(H(Z|U))*\epsilon \overset{\rm (a)}= H_b^{-1}(H(Z|U))(1-2\epsilon) + \epsilon,
\end{align*}
where (a) follows from the definition of the operator-$*$ defined in Table \ref{tab:notation}, which implies that
\begin{align}
    H_b^{-1}(H(Z|U)) \le \frac{\lambda*p - \epsilon}{1-2\epsilon} \le \lambda*p*\epsilon,
\end{align}
where the second inequality follows because the first inequality in \eqref{lemma55} is used,
and thus}
\begin{align}
    H(Z|U) \le H_b(\lambda*p*\epsilon). \label{zubi}
\end{align}
%and {the} binary entropy function is monotonically increasing in the interval $[0,1/2]$.

Likewise, in the direction from $\Tilde{X}$ to $X$, again using Mrs. Gerber’s Lemma, we have that $
    H(X|U) \ge H_b(H_b^{-1}(H(\Tilde{X}|U))*p).
$
This equation implies {that} $\lambda*p \ge H_b^{-1}(H(\Tilde{X}|U))*p$. Since $0 \le p \le 1/2$, it follows that $H_b^{-1}(H(\Tilde{X}|U)) \le \lambda$ or equivalently,
\begin{align}
    H(\Tilde{X}|U) \le H_b(\lambda). \label{tildxubi}
\end{align}

Plugging eqs. \eqref{xubi}, \eqref{zubi}, and \eqref{tildxubi} into {eqs.} \eqref{rsbi}, \eqref{rjbi}, and \eqref{rlbi}, we obtain that
\begin{align}
    R_S &\le H_b(\lambda*p*\epsilon) - (1-q)H_b(\lambda*p) - q, \label{rsbig} \\
    R_J &\ge q + (1-q)H_b(\lambda*p) - H_b(\lambda){,} \label{rjbig} \\
    R_L &\ge 1+q -qH_b(\lambda*p) - {H_b(\epsilon)}. \label{rlbig}
\end{align}
From \eqref{rsbig}--\eqref{rlbig}, {by varying $\lambda$ over the range $[0,\frac{1}{2}]$} and taking the union, the inner and outer bounds on the capacity region match. Hence, the proof of {the} GS model is completed.

\medskip
For the CS model, also fix $\lambda$ such that \eqref{xubi} is satisfied. In the direction from $X$ to $Z$, using {Mrs. Gerber’s Lemma} \cite{wyner} yields that
\begin{align}
H(Z|U) \ge H_b(H_b^{-1}(H(X|U))*\epsilon) = H_b(\lambda*p*\epsilon). \label{zubi22}
\end{align}

Substituting \eqref{tildxubi} and \eqref{zubi22} into \eqref{rjcs}, one can derive that $R_J \ge H_b(\lambda*p*\epsilon) - H(\lambda)$, and by {varying $\lambda \in [0,\frac{1}{2}]$}, the optimal rate region of {the} CS model for binary sources is obtained.
\qed
\section{Proof of Corollary 1}
We give only the case of $\rho^2_2 > \rho^2_3$ as the other case is straightforward.
The proof of the region $\mathcal{R}_G$ (cf.\ \eqref{corollary11}) is shown below.

For the achievability part, fix $0 < \alpha \le 1$. Let $U \sim \mathcal{N}(0,1-\alpha)$ and $\Theta \sim \mathcal{N}(0,\alpha)$. Assume that
\begin{align}
\Tilde{X} = U + \Theta. \label{uxtheta}
\end{align}
Then, we have that
\begin{align}
X &= \rho_1U + \rho_1\Theta + N_x, \label{xutheta} \\
Y &=\rho_1\rho_2U + \rho_1\rho_2\Theta + \rho_2N_x + N_y, \label{yutheta} \\
Z &=\rho_1\rho_3U + \rho_1\rho_3\Theta + \frac{\rho_3}{\rho_2}N_x + N_z.\label{zutheta}
\end{align}
From \eqref{xutheta}--\eqref{zutheta}, we have that
\begin{align}
    I(\Tilde{X};U) &= \frac{1}{2}\log(\frac{1}{\alpha}),~~ I(X;U) = \frac{1}{2}\log(\frac{1}{\alpha\rho^2_1 + 1 - \rho^2_1}), \nonumber \\
    I(Y;U) &= \frac{1}{2}\log(\frac{1}{\alpha\rho^2_1\rho^2_2 + 1 - \rho^2_1\rho^2_2}), \nonumber \\
    I(Z;U) &= \frac{1}{2}\log(\frac{1}{\alpha\rho^2_1\rho^2_3 + 1 - \rho^2_1\rho^2_3}). \label{zu}
\end{align}
Note that due to the Markov chains $U-X-Y-Z$, we can write that
\begin{align}
    I(Y;U|Z) &= I(Y;U) - I(Z;U), \nonumber \\
    I(\Tilde{X};U|Y) &= I(\Tilde{X};U) - I(Y;U). \nonumber \\
    I(X;U|Y) &= I(X;U) - I(Y;U).
\end{align}
Substituting all equations in \eqref{zu} into the right-hand side of \eqref{theorem1} and \eqref{theorem2}, one can see that any rate tuple contained in the right-hand side of \eqref{corollary11} is achievable.

For the converse part, it is a bit more involved. Here, we prove this part by making use of conditional entropy power inequality (EPI) \cite[Lemma II]{bergmans1974}. Note that each constraint in the right-hand side of \eqref{theorem1} can be transformed as
%\vspace{-2mm}
\begin{align}
    R_S &\le I(Y;U|Z) = h(Y|Z) - h(Y|U,Z)  \nonumber \\
    &= \frac{1}{2}\log 2 \pi e (1-\rho^2_3/\rho^2_2)- h(Y|U,Z), \label{iyuzzz} \\
    R_J & \ge I(\Tilde{X};U|Y) \overset{\mathrm{(a)}}= I(\Tilde{X};U|Z) - I(Y;U|Z) \nonumber \\
    &= h(\Tilde{X}|Z) - h(\Tilde{X}|U,Z) - h(Y|Z) + h(Y|U,Z) \nonumber \\
    &= \frac{1}{2}\log\frac{1-\rho^2_1\rho^2_3}{1-\rho^2_3/\rho^2_2} - h(\Tilde{X}|U,Z) + h(Y|U,Z),
\end{align}
\begin{align}
    R_L &\ge I(X;U|Y) + I(Z;X) \nonumber \\
    &\overset{\mathrm{(b)}}= I(X;U|Z) - I(Y;U|Z) + I(Z;X) \nonumber \\
    &\overset{\mathrm{(c)}}\ge \frac{1}{2}\log\frac{1}{1-\rho^2_3/\rho^2_2} - h(X|U,Z) + h(Y|U,Z),\label{ixzuuu}
\end{align}
where (a) and (b) are due to the Markov chain $Z-Y-\Tilde{X}-U$ and $Z-Y-X-U$, respectively, and (c) follows from the property that $h(X)=h(Y)$ for Gaussian RVs with unit variances, and thus $ h(X|Z) - h(Y|Z) + I(Z;X) = I(Z;Y) = \frac{1}{2}\log\frac{1}{1-\rho^2_3/\rho^2_2}$.

So as to bound the region $\mathcal{R}_G$, we need to find a lower bound on $h(Y|U,Z)$ and an upper bound on $h(\Tilde{X}|U,Z)$ for fixed {$h(X|U,Z)$}.

The following fine setting plays {an important} role to bound the conditional entropies $h(Y|U,Z)$ and $h(\Tilde{X}|U,Z)$. Now let us set
\begin{align}
    h(X|U,Z) &= \frac{1}{2} \log 2 \pi e \left(\frac{(\alpha\rho^2_1+1-\rho^2_1)(1-\rho^2_3)}{\alpha\rho^2_1\rho^2_3+1-\rho^2_1\rho^2_3}\right) \label{setting111}
\end{align}
for $0 < \alpha \le 1$. This setting comes from the fact that $\frac{1}{2} \log 2 \pi e \left(\frac{(1-\rho^2_1)(1-\rho^2_3)}{1-\rho^2_1\rho^2_3}\right) = h(X|\Tilde{X},Z) \le  h(X|U,Z) \le h(X|Z) = \frac{1}{2}\log2\pi e(1-\rho^2_3)$. Here, the value $\alpha = 0$ is excluded since it is not achievable for Gaussian RVs with finite variances. This will be mention again later.
%the right hand-side of \eqref{setting111} will go to infinity, but this value is impossible to achieve by Gaussian RVs with finite {variances}, which {are} always assumed in the analysis of Gaussian sources.

In the direction from $X$ to $Y$, using the conditional EPI \cite[Lemma II]{bergmans1974}, we have that
\begin{align}
&e^{2h(Y|U,Z)} \nonumber \\
&\ge e^{2h(\rho_2X|U,Z)} + e^{2h(N_y|U,Z)} \nonumber \\
&~~= \rho^2_2 e^{2h(X|U,Z)} + e^{2h(N_y)} \nonumber \\
&~~\overset{\mathrm{(d)}}= 2 \pi e \rho^2_2 \left(\frac{(\alpha\rho^2_1+1-\rho^2_1)(1-\rho^2_3)}{\alpha\rho^2_1\rho^2_3+1-\rho^2_1\rho^2_3}\right) + 2 \pi e(1-\rho^2_2), \label{hxuz}
%&~~= 2 \pi e \left(\frac{\alpha \rho^2_1(1-\rho^2_2) + (1-\rho^2_1)(\alpha\rho^2_2+1-\rho^2_2)}{\alpha\rho^2_2+1-\rho^2_2}\right),\label{hxuz}
\end{align}
where (c) follows {from} \eqref{setting111}.

Now let us focus only on the numerator of \eqref{hxuz} (inside the biggest parenthesis). We continue scrutinizing it as
\begin{align*}
    &(\alpha\rho^2_1 + 1 - \rho^2_1)(\rho^2_2-\rho^2_2\rho^2_3) + (\alpha\rho^2_1\rho^2_3 + 1 - \rho^2_1\rho^2_3)(1-\rho^2_2) \nonumber \\
    %&=\alpha\rho^2_1\rho^2_2 + \rho^2_2 - \rho^2_1\rho^2_2 -\alpha\rho^2_1\rho^2_2\rho^2_3 -\rho^2_2\rho^2_3 + \rho^2_1\rho^2_2\rho^2_3 + \alpha\rho^2_1\rho^2_3 + 1 - \rho^2_1\rho^2_3 -\alpha\rho^2_1\rho^2_2\rho^2_3 - \rho^2_2 + \rho^2_1\rho^2_2\rho^2_3\nonumber \\
    &=\alpha\rho^2_1\rho^2_2 + 1 - \rho^2_1\rho^2_2 -\rho^2_2\rho^2_3 + \alpha\rho^2_1\rho^2_3 - \rho^2_1\rho^2_3 -2\alpha\rho^2_1\rho^2_2\rho^2_3 \nonumber \\
    &~~~+ 2\rho^2_1\rho^2_2\rho^2_3\nonumber \\
    &=(\alpha\rho^2_1\rho^2_2 + 1 -\rho^2_1\rho^2_2)(1-\frac{\rho^2_3}{\rho^2_2}) + \frac{\rho^2_3}{\rho^2_2} + 2\alpha\rho^2_1\rho^2_3 - 2\rho^2_1\rho^2_3 \nonumber \\
    &~~~- \rho^2_2\rho^2_3 - 2\alpha\rho^2_1\rho^2_2\rho^2_3 + 2\rho^2_1\rho^2_2\rho^2_3 \nonumber \\
    &=(\alpha\rho^2_1\rho^2_2 + 1 -\rho^2_1\rho^2_2)(1-\frac{\rho^2_3}{\rho^2_2}) + \frac{\rho^2_3}{\rho^2_2}(1-\rho^4_2) \nonumber \\
    &~~~+ 2\alpha\rho^2_1\rho^2_3(1-\rho^2_2) - 2\rho^2_1\rho^2_3(1 - \rho^2_2) \nonumber \\
    &=(\alpha\rho^2_1\rho^2_2 + 1 -\rho^2_1\rho^2_2)(1-\frac{\rho^2_3}{\rho^2_2}) \nonumber \\
    &~~~ + \frac{\rho^2_3}{\rho^2_2}(1-\rho^4_2) + 2\rho^2_1\rho^2_3(1-\rho^2_2)(\alpha - 1)
\end{align*}
\begin{align}
    &\ge (\alpha\rho^2_1\rho^2_2 + 1 -\rho^2_1\rho^2_2)(1-\frac{\rho^2_3}{\rho^2_2}) \nonumber \\
    &~~~ + (1-\rho^2_2)(\frac{\rho^2_3}{\rho^2_2}(1+\rho^2_2) - 2\rho^2_1\rho^2_3) \nonumber \\
    &\ge (\alpha\rho^2_1\rho^2_2 + 1 -\rho^2_1\rho^2_2)(1-\frac{\rho^2_3}{\rho^2_2}) \nonumber \\
    &~~~ + (1-\rho^2_2)(\frac{\rho^2_3}{\rho^2_2}(1-\rho^2_1\rho^2_2) + \rho^2_3(1 - \rho^2_1)) \nonumber \\
    &\ge (\alpha\rho^2_1\rho^2_2 + 1 -\rho^2_1\rho^2_2)(1-\frac{\rho^2_3}{\rho^2_2}). \label{numerator1}
\end{align}
where (d) follows because $\alpha \in (0,1]$ and $\rho^2_1,\rho^2_1,\rho^2_3 < 1$.
Plugging \eqref{numerator1} into \eqref{hxuz}, we obtain that
\begin{align}
    e^{2h(Y|U,Z)} &\ge 2 \pi e \left(\frac{(\alpha\rho^2_1\rho^2_2 + 1 -\rho^2_1\rho^2_2)(1-\rho^2_3/\rho^2_2)}{\alpha\rho^2_1\rho^2_3+1-\rho^2_1\rho^2_3}\right).
\end{align}
Therefore,
\begin{align}
    h(Y|U,Z) \ge \frac{1}{2}\log 2 \pi e \left(\frac{(\alpha\rho^2_1\rho^2_2 + 1 -\rho^2_1\rho^2_2)(1-\rho^2_3/\rho^2_2)}{\alpha\rho^2_1\rho^2_3+1-\rho^2_1\rho^2_3}\right). \label{hyuzzzz}
\end{align}

On the contrary, from the direction from $\Tilde{X}$ to $X$, we have that
\begin{align}
    2^{2h(X|U,Z)} &\ge 2^{2h(\rho_1\tilde{X}|U,Z)} + 2^{2h(N_1|U,Z)} \nonumber \\
    2^{2h(X|U,Z)} &\ge \rho^2_1 2^{2h(\tilde{X}|U,Z)} + 2^{2h(N_x)} \nonumber \\
    \rho^2_1 2^{2h(\tilde{X}|U,Z)} &\le 2^{2h(X|U,Z)} - 2^{2h(N_x)}.
\end{align}
It follows that
\begin{align}
    2h&(\Tilde{X}|U,Z) \nonumber \\
    &\le \log \left( \frac{2 \pi e}{\rho^2_1} \left(\frac{(\alpha\rho^2_1+ 1-\rho^2_1)(1-\rho^2_3)}{\alpha\rho^2_1\rho^2_3+1-\rho^2_1\rho^2_3} - (1-\rho^2_1)\right)\right). \label{tilxxuz}
\end{align}
Again let us focus on the numerator of \eqref{tilxxuz}.
\begin{align}
    &(\alpha\rho^2_1 + 1 - \rho^2_1)(1-\rho^2_3) - (1-\rho^2_1)(\alpha\rho^2_1\rho^2_3 + 1 - \rho^2_1\rho^2_3) \nonumber \\
    %&=\alpha\rho^2_1 + 1 - \rho^2_1 - \alpha\rho^2_1\rho^2_3 -\rho^2_3 + \rho^2_1\rho^2_3 - \alpha\rho^2_1\rho^2_3 - 1 + \rho^2_1\rho^2_3 + \alpha\rho^4_1\rho^2_3 + \rho^2_1 - \rho^4_1\rho^2_3 \nonumber \\
    %&=\alpha\rho^2_1 - \alpha\rho^2_1\rho^2_3 -\rho^2_3 + \rho^2_1\rho^2_3 - \alpha\rho^2_1\rho^2_3 + \rho^2_1\rho^2_3 + \alpha\rho^4_1\rho^2_3 - \rho^4_1\rho^2_3 \nonumber \\
    &=\alpha\rho^2_1 - 2\alpha\rho^2_1\rho^2_3 -\rho^2_3 + 2\rho^2_1\rho^2_3 + \alpha\rho^4_1\rho^2_3 - \rho^4_1\rho^2_3 \nonumber \\
    &=\alpha\rho^2_1(1-\rho^2_1\rho^2_3) - 2\alpha\rho^2_1\rho^2_3 -\rho^2_3 + 2\rho^2_1\rho^2_3 + 2\alpha\rho^4_1\rho^2_3 - \rho^4_1\rho^2_3 \nonumber \\
    &=\alpha\rho^2_1(1-\rho^2_1\rho^2_3) - \rho^2_3(2\alpha\rho^2_1 + 1 - 2\rho^2_1 - 2\alpha\rho^4_1 + \rho^4_1) \nonumber \\
    &=\alpha\rho^2_1(1-\rho^2_1\rho^2_3) - \rho^2_3(2\alpha\rho^2_1(1-\rho^2_1) + (1 - \rho^2_1)^2) \nonumber \\
    &\le \alpha\rho^2_1(1-\rho^2_1\rho^2_3).
\end{align}
Therefore, we can derive that
\begin{align}
    h(\Tilde{X}|U,Z) \le \frac{1}{2}\log \left(\frac{2 \pi e\alpha(1-\rho^2_1\rho^2_3)}{\alpha\rho^2_1\rho^2_3+1-\rho^2_1\rho^2_3}\right). \label{htilxuz}
\end{align}

Finally, substituting \eqref{setting111}, \eqref{hyuzzzz}, and \eqref{htilxuz} into \eqref{iyuzzz}--\eqref{ixzuuu}, the converse proof of the region $\mathcal{R}_G$ is completed.

\medskip
For the CS model, we mention only the storage rate since the bounds of the secret-key and privacy-leakage rates follows the arguments of the GS model. 

For the achievability part, sample $U$ into $X$ in the same manners of the achievability proof of the GS model, implying we can still use the relations in Equation \eqref{zu} for the CS model, and observe that
\begin{align}
    I(\Tilde{X};U|Z) = I(\Tilde{X};U) - I(Z;U), \label{ixuzzz}
\end{align}
which is due to the Markov chain $U-X-Z$. Putting the values of $I(X;U), I(Z;U)$ of \eqref{zu} into \eqref{ixuzzz}, we can see that any rate tuple contained in the right-hand side of \eqref{corollary12} is achievable.

Regarding the converse part, fixing an $\alpha$ satisfying \eqref{setting111}, it is straightforward that
\begin{align}
    R_J &\ge I(\Tilde{X};U|Z) = h(\Tilde{X}|Z) - h(\Tilde{X}|U,Z) \nonumber \\
    &\ge \frac{1}{2}\log2 \pi e(1-\rho^2_1\rho^2_3) - \frac{1}{2}\log\left(\frac{2 \pi e\alpha(1-\rho^2_1\rho^2_3)}{\alpha\rho^2_1\rho^2_3+1-\rho^2_1\rho^2_3}\right) \nonumber \\
    &=\frac{1}{2}\log \left(\frac{\alpha\rho^2_1\rho^2_3+1-\rho^2_1\rho^2_3}{\alpha}\right).
\end{align}
Now the needed bound is obtained, and that concludes the proof of converse for $\mathcal{R}_C$ in Corollary \ref{coroll1}.
\qed.

\section{Convexity of the Regions $\mathcal{R}_G$ and $\mathcal{R}_C$}
In this appendix, we show only \eqref{corollary11} because \eqref{corollary12} and \eqref{corollary13} can be proved similarly. It seems difficult to prove this convexity directly via the region expressions in \eqref{corollary11}. Here, we use a technique of changing the parameter $\alpha$ to show the claim. Define
\begin{align}
    \nu = \frac{\alpha\rho^2_1\rho^2_3 + 1 - \rho^2_1\rho^2_3}{\alpha\rho^2_1\rho^2_2 + 1 - \rho^2_1\rho^2_2},
\end{align}
and then it holds that
\begin{align}
    \alpha = \frac{\nu(1-\rho^2_1\rho^2_2)-(1-\rho^2_1\rho^2_3)}{\rho^2_1\rho^2_3-\nu\rho^2_1\rho^2_2}.
\end{align}
Plugging the value of $\alpha$ into the right-hand sides of $R_J$ and $R_L$, we have that
\begin{align}
    R_J &\ge \frac{1}{2}\log\frac{\frac{\nu(1-\rho^2_1\rho^2_2)-(1-\rho^2_1\rho^2_3)}{\rho^2_1\rho^2_3-\nu\rho^2_1\rho^2_2}\cdot\rho^2_1\rho^2_2 + 1 - \rho^2_1\rho^2_2}{\frac{\nu(1-\rho^2_1\rho^2_2)-(1-\rho^2_1\rho^2_3)}{\rho^2_1\rho^2_3-\nu\rho^2_1\rho^2_2}} \nonumber \\
    %&= \frac{1}{2}\log\frac{\rho^2_1(\rho^2_2-\rho^2_3)}{(1-\rho^2_1\rho^2_3)-(1-\rho^2_1\rho^2_2)\nu} \nonumber \\
    &= -\frac{1}{2}\log\left(1-\frac{1-\rho^2_1\rho^2_2}{1-\rho^2_1\rho^2_3}\nu\right) +\frac{1}{2}\log\frac{\rho^2_1(\rho^2_2-\rho^2_3)}{1-\rho^2_1\rho^2_3}.
\end{align}
and
\begin{align}
R_L &\ge \frac{1}{2}\log\frac{\frac{\nu(1-\rho^2_1\rho^2_2)-(1-\rho^2_1\rho^2_3)}{\rho^2_1\rho^2_3-\nu\rho^2_1\rho^2_2}\cdot\rho^2_1\rho^2_2 + 1 - \rho^2_1\rho^2_2}{\left(\frac{\nu(1-\rho^2_1\rho^2_2)-(1-\rho^2_1\rho^2_3)}{\rho^2_1\rho^2_3-\nu\rho^2_1\rho^2_2}\cdot \rho^2_1 + 1 - \rho^2_1\right)(1-\rho^2_1)} \nonumber \\
%&= \frac{1}{2}\log\frac{(\rho^2_2-\rho^3_3)}{((1-\rho^2_3)-(1-\rho^2_2)\nu)(1-\rho^3_1)} \nonumber \\
&= -\frac{1}{2}\log\left(1-\frac{1-\rho^2_2}{1-\rho^2_3}\nu\right) + \frac{1}{2}\log\frac{\rho^2_2-\rho^2_3}{(1-\rho^2_3)^2}.
\end{align}
Since $\nu < 1$ (this is because $\nu$ is an increasing function with respect to $\alpha \in (0,1]$) and $(1-\rho^2_2)/(1-\rho^2_3) < (1-\rho^2_1\rho^2_2)/(1-\rho^2_1\rho^2_3) < 1$, it guarantees that both $1-(1-\rho^2_2)/(1-\rho^2_3)\nu$ and $1-(1-\rho^2_1\rho^2_2)/(1-\rho^2_1\rho^2_3)\nu$ are greater than zero. Therefore, an alternative expression of \eqref{corollary11} is given below.
\begin{align}
    \mathcal{R}_G &=\Big\{(R_S,R_J,R_L) \in \mathbb{R}^+_3: R_S \le \frac{1}{2}\log \nu, \nonumber \\
    &R_J \ge -\frac{1}{2}\log\left(1-\frac{1-\rho^2_1\rho^2_2}{1-\rho^2_1\rho^2_3}\nu\right) +\frac{1}{2}\log\frac{\rho^2_1(\rho^2_2-\rho^2_3)}{1-\rho^2_1\rho^2_3}, \nonumber \\
    &R_L \ge -\frac{1}{2}\log\left(1-\frac{1-\rho^2_2}{1-\rho^2_3}\nu\right) +\frac{1}{2}\log\frac{\rho^2_2-\rho^3_3}{(1-\rho^2_3)^2} \nonumber \\
    &{\rm for}~{\rm all}~\frac{1-\rho^2_1\rho^2_2}{1-\rho^2_1\rho^2_3}<\nu<1\Big\}.
\end{align}

Without loss of generality, suppose that $(1-\rho^2_1\rho^2_2)/(1-\rho^2_1\rho^2_3) \le \nu_1 \le \nu_2 <1$, and the tuples $\boldsymbol{R}_1 = (R^1_S,R^1_J,R^1_L)$ and $\boldsymbol{R}_2 = (R^2_S,R^2_J,R^2_L)$ are achievable for $\nu_1$ and $\nu_2$, respectively. Now obverse that for $0 \le \lambda \le 1$,
\begin{align}
    \lambda R^1_S + (1-\lambda)R^2_S &\le \frac{1}{2}(\lambda\log \nu_1 + (1-\lambda)\log \nu_2) \nonumber \\
    &\overset{\rm (a)}\le \frac{1}{2}\log (\lambda\nu_1 + (1-\lambda)\nu_2) \nonumber \\
    &\overset{\rm (b)}= \frac{1}{2}\log \nu', \label{rsnu}
\end{align}
where (a) follows because $\log x~(x > 0)$ is a concave function and (b) holds since we define $\nu' = \lambda\nu + (1-\lambda)\nu_2$. For the privacy-leakage rate, we have that
\begin{align}
    \lambda &R^1_J + (1-\lambda)R^2_J \nonumber \\
    & \ge  -\frac{\lambda}{2}\log\left(1-\frac{1-\rho^2_1\rho^2_2}{1-\rho^2_1\rho^2_3}\nu_1\right) \nonumber \\
    &~~~- \frac{1-\lambda}{2}\log\left(1-\frac{1-\rho^2_1\rho^2_2}{1-\rho^2_1\rho^2_3}\nu_2\right) + \frac{1}{2}\log\frac{\rho^2_1(\rho^2_2-\rho^2_3)}{1-\rho^2_1\rho^2_3} \nonumber \\
    &\overset{\rm (c)}\ge -\frac{1}{2}\log\left(1-\frac{1-\rho^2_1\rho^2_2}{1-\rho^2_1\rho^2_3}\left(\lambda\nu_1 + (1-\lambda)\nu_2\right)\right) \nonumber \\
    &~~~+ \frac{1}{2}\log\frac{\rho^2_1(\rho^2_2-\rho^2_3)}{1-\rho^2_1\rho^2_3} \nonumber \\
    &= -\frac{1}{2}\log\left(1-\frac{1-\rho^2_1\rho^2_2}{1-\rho^2_1\rho^2_3}\nu'\right) + \frac{1}{2}\log\frac{\rho^2_1(\rho^2_2-\rho^2_3)}{1-\rho^2_1\rho^2_3}, \label{rlnu}
\end{align}
where (c) follows because for real values $0<a,x<1$, the function $-\log(1-ax)$ is concave upward. Similarly, the storage can also be bounded as
\begin{align}
&\lambda R^1_L + (1-\lambda)R^2_L \nonumber \\
&\ge -\frac{1}{2}\log\left(1-\frac{1-\rho^2_2}{1-\rho^2_3}\nu'\right) +\frac{1}{2}\log\frac{\rho^2_2-\rho^3_3}{(1-\rho^2_3)^2}. \label{rjnu}
\end{align}
From \eqref{rsnu}--\eqref{rjnu}, one can see that there exists a $\nu'$ such that $\lambda\boldsymbol{R}_1 + (1-\lambda)\boldsymbol{R}_2 \in \mathcal{R}_G$. This concludes that the region $\mathcal{R}_G$ in \eqref{corollary11} is convex. 


\vfill

\end{document}